\documentclass[%
reprint,
superscriptaddress,
amsmath,amssymb,
aps]{revtex4-2}

\usepackage{graphicx}
\usepackage{bm}

\usepackage[utf8]{inputenc}
\usepackage{placeins}
\usepackage[normalem]{ulem}

\usepackage{color}

\usepackage{multirow} 
\usepackage{makecell} 
\usepackage{booktabs} 
\usepackage{longtable} 
\usepackage{algpseudocode} 
\usepackage{comment} 
\usepackage{hyperref}
\begin{document}
\preprint{APS/123-QED}

\title{Systemic risk mitigation in supply chains through network rewiring} 

\author{Giacomo Zelbi}
\affiliation{Complexity Science Hub Vienna, A-1030 Vienna, Austria}

\author{Leonardo Niccolò Ialongo}
\affiliation{Complexity Science Hub Vienna, A-1030 Vienna, Austria}
\affiliation{Supply Chain Intelligence Institute Austria, A-1030 Vienna, Austria}

\author{Stefan Thurner}
\email{Corresponding author, e-mail: stefan.thurner@meduniwien.ac.at}
\affiliation{Complexity Science Hub Vienna, A-1030 Vienna, Austria}
\affiliation{Supply Chain Intelligence Institute Austria, A-1030 Vienna, Austria}
\affiliation{Section for Science of Complex Systems, CeMSIIS, Medical University of Vienna, A-1090 Vienna, Austria}
\affiliation{Santa Fe Institute, Santa Fe, NM 85701, USA}


\begin{abstract}
The networked nature of supply chains makes them susceptible to systemic risk, where local firm failures can propagate through firm interdependencies that can lead to cascading supply chain disruptions. The systemic risk of supply chains can be quantified and is closely related to the topology and dynamics of supply chain networks (SCN). How different network properties contribute to this risk remains unclear. Here, we ask whether systemic risk can be significantly reduced by strategically rewiring supplier-customer links. In doing so, we understand the role of specific endogenously emerged network structures and to what extent the observed systemic risk is a result of fundamental properties of the dynamical system. We minimize systemic risk through rewiring by employing a method from statistical physics that respects firm-level constraints to production. Analyzing six specific subnetworks of the national SCNs of Ecuador and Hungary, we demonstrate that systemic risk can be considerably mitigated by 16-50\% without reducing the production output of firms. A comparison of network properties before and after rewiring reveals that this risk reduction is achieved by changing the connectivity in non-trivial ways. These results suggest that actual SCN topologies carry unnecessarily high levels of systemic risk. We discuss the possibility of devising policies to reduce systemic risk through minimal, targeted interventions in supply chain networks through market-based incentives.
\end{abstract}

\maketitle

Supply chain networks (SCN) are dynamical socio-economic systems that arise from the spontaneous self-organization of companies transforming sets of inputs into goods and services. In a SCN, nodes represent companies that act as central decision units, determining what to produce, how to produce it, from whom to source inputs, and to whom and at what price to sell outputs. The outcomes of these decisions are temporary supply links that ensure firms have access to the necessary inputs and consumers for their products. A weighted directed link in the SCN, $W_{ij}$, indicates the quantity of a product supplied by firm $j$ to firm $i$. While SCNs are to some extent formed by local optimizing behavior of firms and can be assumed to form locally robust structures with respect to small shocks, they remain vulnerable to larger and systemic supply disturbances. A temporary failure of a company, for example, can trigger disruptions that propagate through the network, amplifying the initial shock to systemically relevant levels and leading to considerable production and delivery disruptions \cite{craighead2007severity, choi2023jit}. Recent and historical events have exposed these fragilities, shaking the assumptions about the stability and reliability of SCNs. SCN disruptions have been studied in the context of natural catastrophes, such as the Thai floods of 2011 \cite{haraguchi2015flood} or the recent earthquakes in Japan \cite{carvalho2020greateast, inoue2019}. More recently, lockdowns and other measures during the COVID-19 pandemic have caused unprecedented SCN disruptions worldwide \cite{pichler2022IOmodel, del2020supply, bonadio2021global} with long-lasting effects on prices and inflation \cite{ascari2024global}. The pandemic caused strong pressures on global supply chains, resulting in, for example, congestion of the U.S. West Coast shipping terminals \cite{kent2022perfect} and shortages in semiconductor chips for the automotive industry \cite{ramani2022understanding}. Since 2022, geopolitical tensions have led to severe shortages in gas and grain supplies from Russia and Ukraine \cite{ben2022impacts,laber2023shock}. In times of changing geopolitical equilibria, the supply of raw materials and technologies across borders is becoming a source of major concern \cite{the_white_house_fact_2024, eu2024bev, draghi2024competitiveness}.

Traditionally, the study of risks related to supply chain dependencies has been constrained by data limitations. Its focus has been on the aggregate sector-level dynamics (input-output analysis) and on focal firms and their direct suppliers and customers –– only sometimes including information on multiple tiers. Recent advances in data availability of firm-level supplier-buyer relations offer unprecedented insights into supply chain dynamics \cite{bacilieri2023whatdowe, pichler2023buildinganalliance}. While traditional Input-Output analysis, which aggregates firms into industrial sectors, fails to capture the heterogeneous nature of company-specific input and output patterns and thus neglects network structures, firm-level data reveal the underlying networks at the scale at which actual business decisions are made \cite{choi2006supply}. This new generation of data demonstrated the importance of detailed network effects in the propagation of production losses by comparing the results of a traditional sector-level scenario with the situation in a real SCN on the firm level \cite{diem2024estimating}.

A key novelty of firm-level data is that they allow the estimation of the consequences of single-firm failures and assess their potential to trigger cascading disruptions \cite{Diem2022ESRI, acemoglu2024macroeconomics}. In other words, it enables the quantification of the systemic risk contribution of individual firms within an economy. Systemic risk emerges from an interplay of the topology of interactions between economic actors, balance sheets, inventories, and the firms' ability to replace their suppliers. Network-based systemic risk has been a subject of academic study in the context of financial markets, in particular, interbank lending markets \cite{boss2004network,iori2006systriskinterbank,battiston2012debtrank, poledna2015multilayer}. Recently, this approach has been adapted and generalized for supply chains \cite{Diem2022ESRI}. There, a firm's systemic importance is quantified by an Economic Systemic Risk Index (ESRI), defined as the fraction of the total production of the economy affected by the failure of that firm. This quantification allows firms to be ranked by their systemic risk contribution (ESRI value), creating ``systemic risk profiles'' of economies. The ESRI profile computed for all Hungarian firms in the national SCN reveals how systemic risk is distributed across the economy and where it is concentrated. Notably, firms with large systemic risk contributions are limited to a tiny subset of businesses and are not necessarily the largest in size or revenue.

In contrast to the static view assumed by most models, the SCN is continuously evolving, with companies rewiring their supply connections and firms entering and exiting the network. Evidence shows that approximately 55\% of all supply relationships present at any given time will disappear within a year \cite{reisch2025}. SCNs reshape their structure in response to shifting economic conditions such as innovations, price changes, and the sudden unavailability of suppliers exiting the market. When a supplier becomes unavailable, a company may respond in many ways, depending on its production function: it can relink to a new supplier, reduce its output, or rely on existing stock inventories. This raises a fundamental question -- how does systemic risk vary with the changing network structure? Is it possible to find solutions within the space of network configurations that are associated with lower systemic risk? And to what extent can systemic risk be mitigated by strategically rewiring supplier-customer links? The substantial level of temporal rewiring of SCNs suggests that changes to the network structure are feasible. If the correct incentives were put in place the system could indeed evolve towards lower systemic risk values with minimal intervention.

In this paper, we explore the potential for systemic risk mitigation in SCN topologies by modifying the network structure of interactions between firms, while maintaining production levels and functions. To this end, we propose a simple link-rewiring algorithm that preserves the production constraints of individual firms, and we apply it to real nationwide firm-level supply chain data. 
In particular, we investigate the extent to which systemic risk can be reduced in six subSCNs, extracted from the countrywide SCNs of Ecuador and Hungary. Our approach is based on the Metropolis-Hastings algorithm, which uses importance sampling in Monte Carlo simulations to choose link modifications that lower systemic risk while avoiding getting stuck in local minima of the systemic risk ``landscape''; see Materials and Methods. A comparison between the systemic risk profiles of the optimal network configurations with those from the original network then allows us to quantify the potential for reducing systemic risk and identify which companies are responsible for this reduction. We then test whether the systemic risk reduction potential depends on specific properties of the network and firm.

The idea of finding networks with minimal systemic risk levels has been explored in the context of financial networks. Previous studies have examined the effect of taxes on interbank links that create systemic risk, demonstrating a significant mitigation potential of up to 50\% \cite{Poledna2016SRtax, leduc2017incentivize}. In a different approach, interbank lending networks were optimized by employing a Mixed-Integer Linear Programming algorithm with the constraint that banks' total assets and liabilities remained untouched. Again, the rewired interbank networks showed a significant reduction of systemic risk of about 70\% \cite{diem2020minimal}. Another work focused on portfolio optimization to reduce systemic risk arising from overlapping bond exposures in European banking stress-test data \cite{Pichler2021nwoptimization}. However, these approaches cannot be directly applied to supply chain data, as supply interactions between firms differ fundamentally from financial exposure. Unlike financial networks, SCNs must adhere to production constraints, geographical limitations, and technological compatibility requirements, introducing additional layers of dynamics.

\begin{figure*}[t!]
    \centering
    \includegraphics[width=510pt]{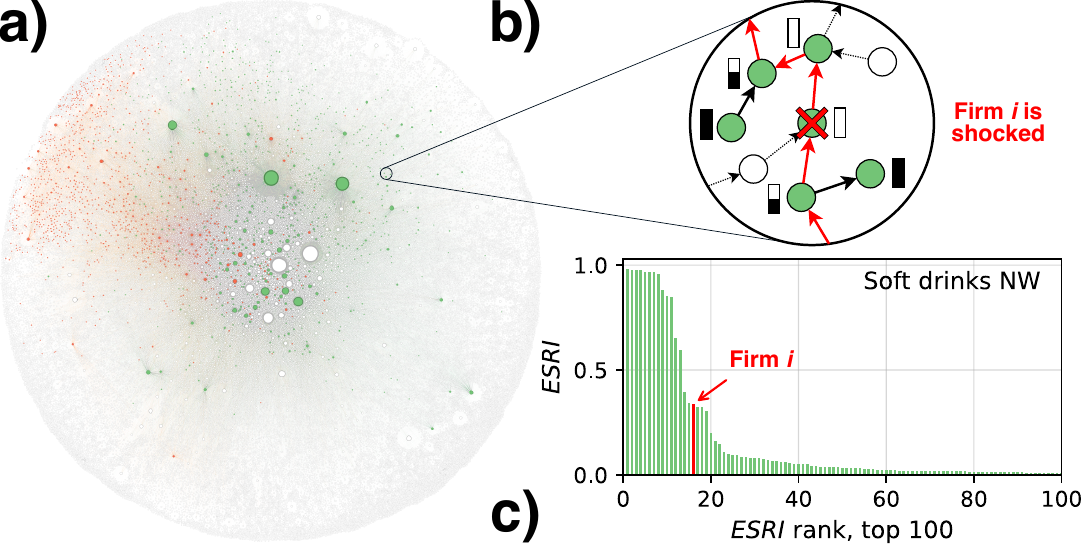}
    \caption{(a) Ecuadorian supply chain network comprised of 65'614 firms and 650'931 supply links. The crustaceans and soft drinks subnetworks are highlighted by red and green nodes. 
    (b) Schematic view of a shock propagation event along the SCN after the default of a particular firm $i$. Suppliers and customers of firm $i$ have to adjust and reduce their production (full bars indicate 100\% production level). Note that the shock propagates up and downstream the SCN affecting the firms' production levels represented by the bars. Panel c) Systemic risk profile. The panel shows the rank distribution of the systemic risk contributions of all firms (ESRI value) in the soft drinks SCN.}
    \label{fig:supply chain shock}
\end{figure*}

To evaluate the systemic risk content of a given SCN, we use the Economic Systemic Risk Index (ESRI) as proposed in \cite{Diem2022ESRI}. The ESRI of firm $i$ quantifies the proportion of the total production lost due to demand and supply disruptions propagating up and down through the SCN, following the failure of firm $i$. An example of such a cascade is shown in Fig.~\ref{fig:supply chain shock} \textbf{b)}, where we zoom in on the neighborhood of a firm, $i$, in the Ecuadorian national SCN depicted in \textbf{a)}. The disruption propagates upstream and downstream, via the directed links connected to $i$. The firms (represented as circles) that depend on $i$ for inputs experience shortages and scale down their production, while suppliers to $i$ reduce their production in response to the loss of a customer. The cascade propagates and affects second-tier firms and so on. ESRI$_i$ is the fraction of total production remaining after the shock has spread compared to the total production before the default of $i$. Note that ESRI takes into account the heterogeneous nature of production functions (firms can have essential or non-essential inputs), as well as the replaceability of suppliers. Computing ESRI for all firms in the system yields the risk profile of the SCN, as shown in Fig.~\ref{fig:supply chain shock} \textbf{c)}. To quantify the overall systemic risk of the SCN, we use the average ESRI across all firms in the SCN, $\langle \text{ESRI}\rangle$. For further details on ESRI, see Materials and Methods, SI section~\ref{SI:sectESRI} and \cite{Diem2022ESRI}. 

\begin{figure}[t!]
\centering
\includegraphics[width=240pt]{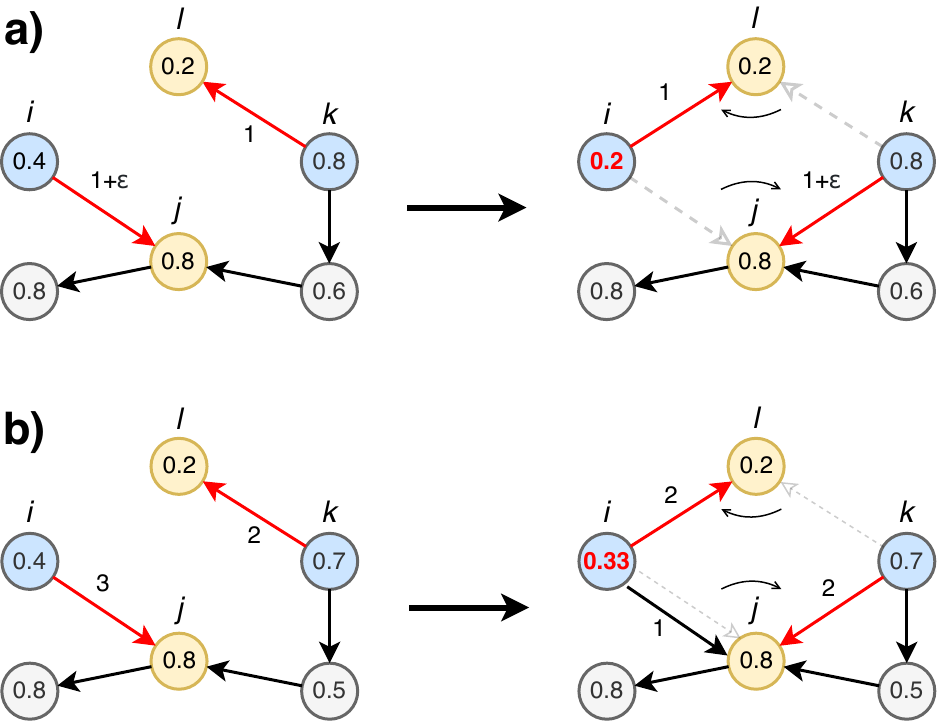}
    \caption{Schematic example of a constrained network rewiring step. Node color represents the economic activity (a proxy for the goods manufactured or services provided) and the numbers inside the nodes indicate their contribution to systemic risk, ESRI. At every rewiring iteration, two directed supply links are randomly selected, ensuring they share the same ordered combination of economic activities at their source and target nodes. In this example (left side of panels), the selected links (red arrows) are $i\to j$ and $k\to l$, both connecting blue nodes to yellow nodes. (a) If the two links have similar weights within a set tolerance threshold, the rewiring step (fat black arrow) swaps the suppliers of the target nodes $j$ and $l$. As a result, $i$ now sells to $l$, and $k$ sells to $j$ (right side). A constraint prevents small weight differences from accumulating over multiple swaps. (b) If the difference in the link weights exceeds the tolerance threshold, the link with the larger weight is split, and only the portion matching the smaller weight is swapped. After rewiring, the topology of the production network changes, and so do the failure cascades following a local shock, and accordingly ESRI values for some nodes will change. Here node $i$ now has a decreased ESRI; the difference is only due to the change in topology.
    }
    \label{fig:SWAP_example}
\end{figure}

\begin{figure*}[t!]
\centering
\includegraphics[width=510pt]{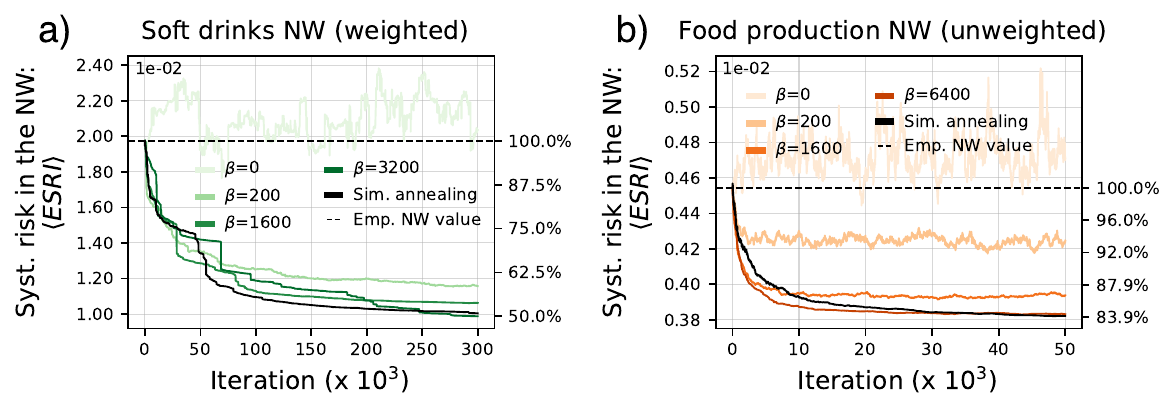}
    \caption{Decrease of systemic risk, $\langle\text{ESRI}\rangle$, as a function of rewiring (Monte Carlo) steps for (a) the soft drinks subnetwork of Ecuador (2015) and (b) the food production subnetwork of Hungary (2017), the latter representing an unweighted SCN. Results are shown for different values of the ``inverse temperature'', $\beta$. For $\beta=0$, the ``configuration model'', every relinking step is accepted and, interestingly, the systemic risk level remains close to the empirical value. For larger $\beta$, the system quickly converges to plateaus of lower-risk network configurations, with the final risk level depending on $\beta$. While higher $\beta$ generally leads to lower and greedier minima convergence, it also increases the likelihood of getting trapped in local ones. This is particularly evident for $\beta=3200$ in (a), where several times plateaus form, although occasionally a lucky swap explores a new local minimum. In the unweighted case (b) this effect is less pronounced. To ameliorate this problem, we employ the simulated annealing method (black lines). The empirical systemic risk values (initial level) are shown as the dashed line.
    The simulations are stopped after 300,000 and 50,000 iterations for the soft drinks and food production networks, respectively.
    }
\label{fig:TotESRI_MCtrajectories}
\end{figure*}

{\it Constraints for the rewiring algorithm.}
To estimate the potential for systemic risk mitigation through rewiring, it is necessary to determine which network configurations are allowed. Not all possible connections between firms are meaningful or feasible -- most are not. The objective of our algorithm is to maintain the system in its current state except for a number of firms changing suppliers. We therefore do not change the production technology of each firm, nor its capacity. This means preserving the empirical technical coefficient given by the ratio of intermediate input per unit of output. We do so by making sure that the total output and input per product of each firm remain close to its original value. In practice, we allow for a maximum change in total production of 20\%, while keeping the input quantities per product exact. In addition, since connections are costly to establish and given the relation between a firm's connectivity and its systemic importance (see SI section~\ref{SI:sectSCATTERPLOTS}), we do not want to significantly change the number of suppliers the firm has for each of its intermediate inputs. To guarantee this, we employ an algorithm based on link swapping that ensures that a supplier is replaced by another firm providing the same product or service. For further details on the rewiring steps, see SI section~\ref{SI:sectREWIRINGRULE}.

To explore the space of supply chains that satisfy these constraints, we develop a Monte Carlo link-swapping algorithm. A schematic representation is shown in Fig.~\ref{fig:SWAP_example}. At every time step, we randomly select two supply links (red arrows, $i \to j$ and $k \to l$) that connect companies with the same pair of economic activities, indicated by colors in the figure (e.g., blue ships to yellow, with matching colors at the source and target). If the two links have the same weight (as in panel (a)), we swap them entirely. This swap (fat black arrow) can be interpreted as exchanging the suppliers of the target nodes or the customers of the source nodes. However, since supply link weights represent product volumes, it is unlikely that two links will have exactly the same weight. To address this problem, we introduce a small tolerance by allowing full swaps when weights differ by a value of $\epsilon$, or less. To prevent unintended accumulation of changes, we impose a constraint that ensures that a firm’s total out-strength does not deviate by more than 20\% from its original value. Note that this swap does not change the overall degree distributions in the SCN. If instead the difference in link weights is too large for a full swap, the link with the larger weight is split, and only a portion equal to the smaller weight is exchanged (as shown in panel (b)). This introduces a new link but preserves the technical coefficients of all firms involved. If no weight information is available, all links can be considered of the same weight, and the algorithm operates as in case (a). All swaps are reversible, ensuring free exploration of configuration space.

Even under these strict constraints, for realistic SCN sizes, the number of possible network configurations is immense, and computing the average systemic risk of every configuration is not feasible. Therefore, we sample them using a Metropolis-Hastings algorithm with an acceptance criterion that biases the link-rewiring moves toward lower-risk configurations. After every swap iteration, we compute the average ESRI for the new trial SCN. If the $\langle\text{ESRI}\rangle$ in the new configuration is lower, the swap is accepted; otherwise, if the trial configuration leads to a higher $\langle\text{ESRI}\rangle$, the swap is accepted only with a probability $p = \exp(-\beta (\langle\text{ESRI}\rangle_{after}-\langle\text{ESRI}\rangle_{before}))$. If the swap is rejected, we continue with the next swap iteration. The parameter $\beta > 0$ is sometimes called an ``inverse temperature'', and it controls the likelihood of accepting higher $\langle\text{ESRI}\rangle$ moves, helping the system to escape local minima. By gradually increasing $\beta$ over swap iterations, the system converges toward an optimal configuration (simulated annealing). To calibrate these increasing $\beta$ curves, we first run simulations at fixed $\beta$. See Materials and Methods for more details.

\section*{Results}
We apply the rewiring algorithm to six empirical subnetworks to assess its impact on network resilience, across different production networks. From Ecuador's 2015 SCN, we analyze two food industry subnetworks: the processing of fish, crustaceans and molluscs network and the manufacturing of soft drinks network. From Hungary's 2017 SCN, we extract the unweighted networks for food production and the automotive industry. Each subnetwork comprises approximately 1,000 nodes and fewer than 10,000 supply links, making $\langle \text{ESRI} \rangle$ computation feasible at every swap iteration. For details on subnetwork extraction, see SI section~\ref{SI:extractECU} and~\ref{SI:extractHUN}. In the VAT datasets there is no information on the products being exchanged. Therefore, we choose the NACE 3-digit classification \cite{naceclassification2006} of the supplying firms as a proxy for the product classification of the link used in the computation of ESRI. Note that the swapping algorithm not only guarantees that the amount of inputs by product used by each firm does not change, but also guarantees that the output by receiving industry is not modified. This further guarantees that the input-output structure of the network is not altered. 

We iteratively rewire each subnetwork, starting from the true empirical configuration. 
Figure~\ref{fig:TotESRI_MCtrajectories} shows the average $\langle\text{ESRI}\rangle$ as a function of rewiring steps, for (a) the soft drinks network and the (b) unweighted food production network. Colored curves represent different fixed-$\beta$ rewiring simulations, and the black lines correspond to simulated annealing ($\beta$-increasing). The dashed line indicates the systemic risk level of the empirical subnetworks.
For $\beta=0$ (accepting every swap regardless of risk impact), the systemic risk remains around the empirical level, indicating that without intervention, SCNs tend to settle in suboptimal configurations far from minimal systemic risk. The $\beta=0$ case is sometimes referred to as  ``configuration model''.

For higher $\beta$ values, the simulations converge to different plateaus of low-risk network configurations after several tens of thousands of rewiring steps. Notably, 10,000 steps roughly correspond to one update for each supply link, equivalent to less than two years of real-world rewiring rates observed in the Hungarian SCN \cite{reisch2025}. The choice of (fixed) $\beta$ seems to select the systemic risk level at convergence, with higher values generally leading to lower risk. However, excessively high $\beta$ increases the chance of getting trapped in local minima. In some cases, the algorithm still finds a move that allows the system to escape, triggering a sharp drop in systemic risk, as seen for $\beta=3200$ in Fig.~\ref{fig:TotESRI_MCtrajectories} \textbf{a)}. To find a balance between finding lower plateaus and the risk of getting stuck, simulated annealing varies $\beta$ throughout the simulation, (see Methods), typically yielding the best results. While these results are likely not global minima, they provide a reasonable estimate of the achievable level of risk reduction. For additional $\beta$ values and other networks, see SI section~\ref{SI:sectALLBETA}. In summary, the systemic risk reduction varies across the networks, ranging from 16\% to 50\%, as summarized in Table~\ref{tab:summary-table} for the different production networks.

\begin{figure*}[!t]
\centering
\includegraphics[width=510pt]{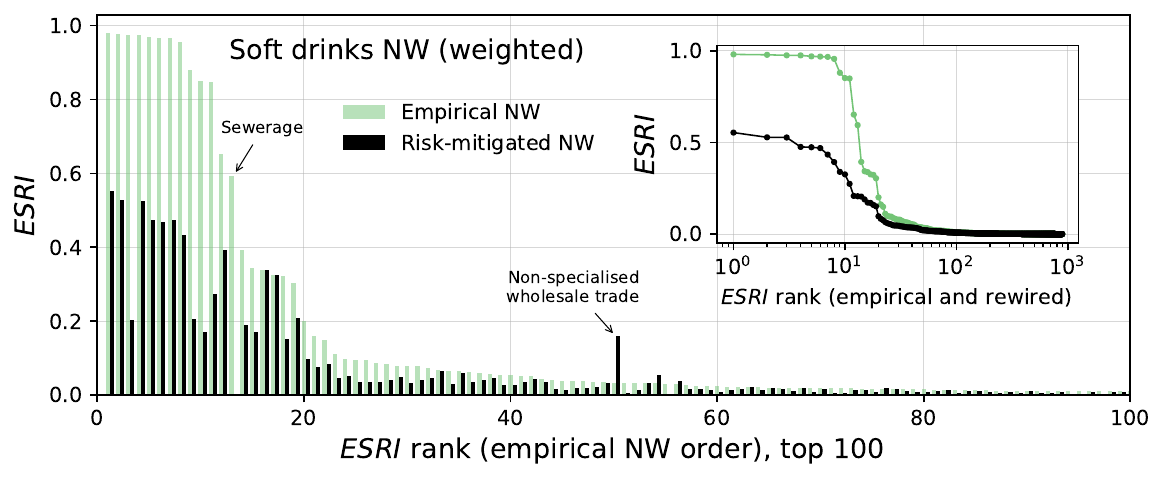}
    \caption{Systemic risk profile of the empirical (light green bars) and risk-mitigated (black bars) soft drinks network. Firms are ranked by their ESRI values in the empirical network along the $x$-axis, displaying only the top 100 riskiest firms. Rewiring reduces the systemic risk contribution of most firms, with a few exceptions where risk increases (as for a firm in Non-specialised wholesale trade). The largest reductions occur among the highest-risk firms, while the systemic risk contribution of the distribution's tail does not change significantly after rewiring.
    The inset presents the same profiles, with the rewired profile now rank-ordered for direct comparison.
    }
    \label{fig:ESRI_bar_profiles_hor}
\end{figure*}

We find that risk mitigation through relinking is nearly as effective for unweighted networks, Fig.~\ref{fig:TotESRI_MCtrajectories} \textbf{b)}, suggesting that network topology is at the heart of systemic risk, while link weights only play a secondary, though important, role. We also note that the convergence appears to be substantially faster for unweighted networks, likely because random swaps frequently target links of negligible importance -- of which there are many due to the fat-tailed distribution of link weights. 
To further examine the role of weights, we constructed unweighted versions of the Ecuadorian soft drinks and crustaceans networks and compared the results of the rewiring algorithm (Table~\ref{tab:summary-table}). The initial average ESRI values for the unweighted case closely match those of the crustaceans network but differ significantly for the soft drinks network, where the empirical ESRI values are nearly twice as high in the weighted version. This could be due to several factors, one hypothesis being the substitutability of firms that depends on the market share of the firm in a given product. The weighted information significantly alters this market share and can therefore lead to the increase of systemic relevance of firms. 
In terms of risk mitigation potential, the crustaceans network again shows similar results between weighted and unweighted cases, while for the soft drinks network, the weighted version achieves higher risk reduction. A comparison of the risk profiles of the weighted and unweighted crustaceans subnetworks (see SI section~\ref{SI:sectBESTPROFILES}), yields that in the unweighted case the risk is concentrated within a very small set of firms, just 5 compared to the 11 in the weighted case. It seems that when the risk is so accumulated in a few firms, the algorithm cannot find possibilities for significant risk reduction. Note that also in the other case of a smaller reduction, the Hungarian food subnetwork, the risk is mainly due to 2 firms, which suggests that in these cases the production constraints do not allow further mitigation.

Rewiring severely affects the systemic risk profile, i.e., the rank-ordered distribution of firms, as shown in Fig.~\ref{fig:ESRI_bar_profiles_hor}. The top 100 riskiest firms in the Ecuadorian soft drinks production network are compared before (light green bars) and after applying the rewiring algorithm with simulated annealing (black bars). The ESRI values of the riskiest firms shrink considerably -- often by 50\% or more. For one firm (in NACE 37.0 - sewerage), it drops to almost zero. Only a few firms firm increase their ESRI after rewiring, as the one indicated by the arrow (NACE 46.9 - Non-specialized wholesale trade). The inset shows the same systemic risk profiles, now with the rewired profile also rank-ordered. Similar results for other networks are shown in SI section~\ref{SI:sectBESTPROFILES}.

To find out how rewiring alters the structural properties of SCNs beyond ESRI values, we compare several network metrics before and after rewiring in Table~\ref{tab:summary-table}. It is evident that there are only a few notable changes to these measures. As expected, the weighted rewiring tends to introduce new links rather than aggregate them. Interestingly, the increase in the number of links, $L$, does not correlate with a reduction in ESRI, but rather appears to be a consequence of how the configuration space is explored. Indeed, ``configuration model'' simulations (unbiased rewiring at $\beta=0$), significantly increase $L$ without changing $\langle\text{ESRI}\rangle$, thus showing that a high $L$ is not a sufficient condition for risk mitigation. In SI section~\ref{SI:sectBETA0}, we show in detail the trajectories of $L$, and we summarize the results in the table.

\begin{table*}[t]
    \centering
    \resizebox{500pt}{!}{%
    \begin{tabular}{|rl|rrrrrrrrrrrr|}
        \toprule
        & & $\langle\text{ESRI}\rangle$ & \thead{$\langle\text{ESRI}\rangle$\\ reduction} & N & L & $\langle k_{tot}\rangle$ & $\langle \langle k_{tot}\rangle_{\text{NN}} \rangle$ & \thead{Global\\clustering\\coefficient}& Diameter$^\star$ & \thead{Average\\ shortest\\path}$^\star$ & \thead{Size of the 3\\largest SCCs} & \thead{Size of the \\ largest WCC} & Reciprocity\\
        \midrule
        \multicolumn{2}{|c|}{\textbf{Weighted NWs (Ecuador)}} &&&&&&&&&&&& \\[5pt]
        Crustaceans     & Empirical      & 0.794E-02                     & - & 1075 & 7128  & 13.26 & 15.15 & 2.6E-05  & 7    & 2.91 & [736,2,2] & 1075   & 12.0\%\\
                        & Risk-mitigated & 0.458E-02 & -42.4\% & -    & 22265 & 41.42 & 33.29 & 0.7E-05  & 5    & 2.25 & [760,2,1] & 1075   & 7.4\% \\
                        & Configuration model ($\beta=0$) & 0.808E-02  & +1.8\% & -    & 21620 & 40.22 & 33.68 & 0.7E-05  & 5.05 & 2.26 & [761,2,1] & 1075.0 & 7.6\% \\ [5pt]
        Soft drinks     & Empirical      & 1.975E-02 & - & 890  & 9323  & 20.95 & 45.16 & 2.3E-05  & 6    & 2.38 & [545,2,2] & 890    & 14.8\%\\
                        & Risk-mitigated & 0.988E-02 & -50.0\% & -    & 13177 & 29.61 & 39.08 & 0.9E-05  & 5    & 2.26 & [542,1,1] & 890    & 8.7\% \\
                        & Configuration model ($\beta=0$) & 2.154E-02 & +9.1\%  & -    & 13329 & 29.95 & 39.32 & 0.9E-05  & 5.10 & 2.27 & [568,1,1] & 890.0  & 9.2\% \\ [5pt]
        \midrule
        \multicolumn{2}{|c|}{\textbf{Unweighted NWs (Ecuador)}} &&&&&&&&&&&& \\[5pt]
        Crustaceans     & Empirical      & 0.804E-02 & - & 1075 & 7128  & 13.26 & 15.15 & 2.35E-01 & 7    & 2.91 & [736,2,2] & 1075   & 12.0\%\\
                        & Risk-mitigated & 0.452E-02 & -43.7\% & -    & -     & -     & 16.65 & 1.51E-01 & 5    & 2.76 & [737,2,1] & 1075   & 7.7\% \\
                        & Configuration model ($\beta=0$) & 0.726E-02 & -9.7\% & -    & -     & -     & 15.58 & 1.88E-01 & 6.1  & 2.76 & [749,2,1] & 1075.0   & 3.4\% \\ [5pt]
        Soft drinks     & Empirical      & 1.142E-02 & - & 890  & 9323  & 20.95 & 45.16 & 3.08E-01 & 6    & 2.38 & [545,2,2] & 890    & 14.8\%\\
                        & Risk-mitigated & 0.933E-02 & -18.3\%  & -    & -     & -     & 49.36 & 2.95E-01 & 5    & 2.32 & [542,2,1] & 890    & 10.4\%\\
                        & Configuration model ($\beta=0$) & 1.220E-02 & +6.8\%  & -    & -     & -     & 45.23 & 3.21E-01 & 5.48 & 2.34 & [565,1,1] & 889.7  & 10.4\%\\ [5pt]
        \midrule
        \multicolumn{2}{|c|}{\textbf{Unweighted NWs (Hungary)}} &&&&&&&&&&&& \\[5pt]                        
        Food production & Empirical & 0.455E-02 & - & 1062 & 2342  & 4.41  & 7.33  & 6.47E-02 & 9    & 3.91 & [127,14,6]& 1062   & 8.5\% \\
                        & Risk-mitigated & 0.381E-02 & -16.1\% & -    & -     & -     & 7.55  & 5.63E-02 & 9    & 3.86 & [62,6,5]  & 1062   & 5.6\% \\
                        & Configuration model ($\beta=0$)   & 0.472E-02 & +3.7\% & -    & -     & -     & 7.37  & 5.13E-02 & 9.71 & 3.86 & [194,5,3] & 1054.6 & 2.8\% \\ [5pt]
        Automotive      & Empirical      & 0.550E-02 & - & 1147 & 2561  & 4.47  & 4.37  & 8.97E-02 & 11   & 4.15 & [237,3,3] & 1147   & 12.8\%\\
                        & Risk-mitigated  & 0.367E-02 & -33.3\%& -    & -     & -     & 4.36  & 4.93E-02 & 10   & 4.05 & [128,3,3] & 1147   & 7.7\% \\
                        & Configuration model ($\beta=0$) & 0.590E-02 & +7.3\% & -    & -     & -     & 4.25  & 4.60E-02 & 10.17& 4.05 & [279,4,3] & 1133.7 & 5.7\% \\ [5pt]

        \bottomrule
    \end{tabular}
    }
    \caption{Risk mitigation potential and network proprieties of six supply chain networks, before and after rewiring. The first column shows the average ESRI values for each empirical, risk-mitigated, and configuration model (unbiased rewiring at $\beta=0$) network. The second column reports the risk reduction in \%. 
    To see which topological changes lead to the observed systemic risk mitigation, we compare several network measures before and after rewiring. For reference, we include the values of the configuration model ($\beta=0$), taking their average value in the random network exploration after the burn-in steps. 
    Measures marked by $\star$ are computed in the respective undirected network, where multi-edges were collapsed to a single link. We find no particular topological measure that explains systemic risk mitigation. This supports the hypothesis that systemic risk emerges from meso-scale structures that cannot be captured by network local measures or global network metrics.
    }
    \label{tab:summary-table}
\end{table*}

Another evident pattern is that reciprocity is substantially reduced in the risk-mitigated network. Reciprocity is overexpressed in real production networks with respect to a suitably defined configuration model \cite{bacilieri2023whatdowe}, and it is clear that our rewiring algorithm, which does not constrain reciprocity levels, will necessarily lower them. This is confirmed by the simulations with $\beta=0$, see SI section~\ref{SI:sectBETA0}. We find that rewiring the unweighted networks reduces the sizes of the largest strongly connected component (SCC) in the Hungarian networks but not in the Ecuadorian ones. Interestingly, the largest SCCs in the Hungarian networks are small (less than 21\%). Given that we do not observe a similar shrinking of the SCC for the other networks, this can be ruled out as a driving factor for systemic risk mitigation. In general, we could not find a pattern between any change in network measures, such as clustering, diameter, or average path length, and the systemic risk mitigation. In SI section~\ref{SI:sectdeltaKdeltaESRI}, we checked whether an increase in a firm's degree leads to higher ESRI, as suggested by their correlation, or instead, reduces it due to the diversification effect.

The lower risk mitigation potential in the Hungarian sub-SCNs may result from the two different ways of constructing the subnetworks. While for Ecuador we first select firms in a target sector and then add all firms in the first tier of suppliers and customers, in the case of Hungary we used a clustering algorithm to identify communities of strongly interconnected nodes; see SI section~\ref{SI:extractHUN}. It is clear that risk reduction depends on the properties of the subnetwork. In this case, the lower mitigation potential might be associated with the lower average degree in the Hungarian networks as well as the considerably smaller size of the largest SCC (see Tab.~\ref{tab:summary-table}).
Except for these properties, we find no particular network measure that is directly associated with systemic risk mitigation. This suggests that systemic risk depends on the network structure at a meso scale as explored in \cite{Diem2022ESRI} that is not captured by the commonly used measures.

\section*{Discussion}

Traditional approaches to increase the resilience of supply chains typically adopt a firm-level perspective, emphasizing different strategies available to firms including higher inventory levels or supplier redundancy \cite{ho2015supply}. Both strategies come at the expense of efficiency and can entail significant costs. While economic literature has largely focused on the debate on reshoring global value chains (GVCs) \cite{baldwin2022risks, König2022PNAS} and maintaining strategic stockpiles for critical industries \cite{sprecher2015framework}, these measures do not fully address systemic risk in supply networks. Similarly, management science has a well-established tradition of examining resilience, primarily through the lens of the flexibility-versus-redundancy debate \cite{kamalahmadi2016review}. However, the role of network topology in mitigating supply chain disruptions has been largely overlooked. Here, we propose an alternative approach that leverages topological considerations to reduce systemic risk. Rewiring supply chain networks (SCNs) does not necessarily impose additional costs or compromise efficiency. Empirical studies indicate that firms adjust supply links at a surprisingly rapid pace \cite{reisch2025}, suggesting that the associated costs may be relatively low. We focus on targeted restructuring of SCNs to limit the spread of local disruptions and prevent cascading failures across the broader economy, thereby mitigating costly supply chain crises. Our findings demonstrate that such rewiring can achieve up to a 50\% reduction in systemic risk without diminishing firms' production capacity.

In particular, we find a risk mitigation potential of -42\% and -50\% for the Ecuadorian ``crustacean'' and ``soft drinks'' production networks, respectively. Topological risk mitigation is even possible for unweighted networks which underscores the essential role of topology in economic systemic risk. We show that similar percentages can be reached in situations, where only the topology of supply relations is known but transaction volumes are not. For this, we produce unweighted versions of the SCNs and find a risk mitigation potential of -44\% and -19\%, respectively. We demonstrate that the risk mitigation margins vary with the specific production network under study. We analyzed different production sectors in food production in two countries. We find an overall range of topological risk reduction between -16\% and -50\%. Our results reinforce the existence of large margins of topological systemic risk reduction, aligning with the levels observed in various financial market contexts, such as interbank markets \cite{Thurner2013transparency, Poledna2016SRtax, diem2020minimal}, bank stress testing of overlapping portfolios \cite{Pichler2021nwoptimization}, and multi-layer exposures \cite{poledna2015multilayer}.

To understand the mechanisms of topological systemic risk mitigation, we correlate the risk reduction with the corresponding changes of a series of local network measures, such as degree, clustering, reciprocity, and global parameters, such as diameter, and size of strongly and weakly connected components. We find no conclusive evidence where one network measure alone could explain the risk reduction. We relate this to the existence of a ``systemic risk core'' in SCNs, as reported in \cite{Diem2022ESRI}. Such a core consists of those firms that carry the bulk of the systemic risk. The existence of a core is indicated by a plateau of high systemic risk firms in the systemic risk profile; as can be seen in the ``Empirical NW'' profiles in Fig.~\ref{fig:ESRI_bar_profiles_hor}. Many of the firms within the core are mutually connected and the failure of one of them typically leads to a failure of most firms in the core. This also explains the formation of the plateau, since any failure would lead to the same set of collapsing firms (creating the same total losses reflected in ESRI). The core is neither characterized by local network properties nor global ones -- it is a network structure at a meso level. Any reconfiguration of the core has a large impact on the total systemic risk in the system. Since this meso-structure is not captured by traditional network measures, and it is not trivial to reduce to a single value, it is reasonable that we cannot attribute the observed systemic risk reduction to one of the network measures alone.

The existence of a risk mitigation potential in the observed six examples means that actual production networks are suboptimal with respect to systemic risk. This is somewhat expected since firms do not and cannot consider systemic risk for decisions on how to choose their suppliers. We find that the levels of systemic risk observed in actual production networks are roughly the same as those that we get with random rewiring with the mentioned constraints. Only when we impose the risk reduction bias in the rewiring, we reach lower levels. This means that if one could make systemic risk information visible to firms, and if this risk could become a factor in their choice of suppliers through the internalization of this cost, market forces would bring the system toward lower systemic risk levels.  We acknowledge that our data do not allow us to investigate the welfare costs of the proposed intervention in terms of loss of productivity or increases in consumer prices. Further, there are many considerations that go into the selection of suppliers, for example, the quality of their products \cite{demir2024ring}, that might further constrain or introduce costs to our proposed method. We have nevertheless established the potential effectiveness of this structural approach to risk mitigation and leave further analysis of the implication and cost-benefit analysis to future studies.

An obvious limitation of the constraints used to generate the space of allowed networks is that we do not have explicit information on the products the firms actually exchange. We proxy firms' products by the NACE classifications of the selling firm that can introduce errors in the computation of their systemic risk. When it comes to estimating the rewiring options of firms -- the sets of potential suppliers for firms -- instead we use as constraints both the target and source node industry classification. This further restriction should ensure that we are more likely to be underestimating the potential systemic risk reduction than overestimating it. Unfortunately, presently there is a lack of data on product information for firm-level networks, making the estimation of the sensitivity of our result to this classification a daunting task. Indeed, the error could go in both directions, the method could be too stringent, by limiting possible swaps to firms in sector pairs where we observe links, or it could be too lax, by assuming that firms in the same sector are mutually substitutable, where, in reality, they specialize in very different production. A further data limitation is that we do not have information on the final demand, meaning that the true size of every firm in the network could have been wrongly assessed. Only the availability of more detailed data will make it possible to improve these shortcomings. 
In a similar direction, a further limitation due to the lack of product information arises with the definition of systemic risk itself. The computation of ESRI involves estimates of the ``substitutability'' of firms in the SCN and their production functions. The current approach involves an approximation and is known to be subject to further improvement until data with more detailed product information becomes available. These issues are well known and were discussed in the literature before \cite{Diem2022ESRI}.

The presented rewiring algorithm could be immediately improved by considering the geographic distance between companies and its influence on establishing supplier-buyer relations \cite{bernard2019production}. The preference of firms for ties to nearby businesses could be added as a further constraint to the rewiring algorithm or as a bias in the probability of accepting the swap. Doing so could help the model retain the well-documented tendency of firms to co-locate for the positive externalities this entails, especially in terms of labor pooling and knowledge transfers \cite{juhasz2024colocation}.

A challenge that is intrinsic to the presented approach is the size of the ``configuration space'' of SCNs. The number of possible networks producing more or less the same set of intermediary and final products is immense. The chances that a rewiring procedure finds the globally optimal (minimal systemic risk) network configurations is practically zero. Therefore, here we do not search for the optimal network configurations, we only make statements of how feasible it is to reach configurations that are significantly less risky. The situation is similar to many problems in statistical physics with large configuration spaces. There, the method of importance sampling has been applied with great success. Its strength is its ability to explore large configuration spaces effectively to find regions of low systemic risk. The problem of getting stuck in local minima can be ameliorated by superimposing methods like simulated annealing, which we successfully implemented here. In our simulations, we find that after several thousand rewiring iterations one converges towards relatively stable risk levels -- with no guarantee of being close to the true optimum. 

Depending on whether we use weighted or unweighted networks to reach approximate convergence one needs in the order of 100,000 and 10,000 rewiring steps, respectively. To find what these numbers mean in the context of actual SCN rewiring one has to consider the number of links in the networks. This is approximately 10 steps per link in the weighted case (soft drinks), and less than one step per link in the unweighted case (food production). Interestingly, in the unweighted case, this roughly corresponds to the turnover of suppliers we would observe in 2-3 years \cite{reisch2025}. In the weighted case currently our algorithm is not as efficient, however in both cases we are considering the steps necessary for the exploration of the configuration space rather than the minimal set of swaps necessary to reduce the systemic risk significantly. It is quite likely that a very small set of link changes is responsible for the majority of the mitigation we observe. Identifying these links as well as the optimal steps necessary to achieve this reduction is the subject of further study. 

Given the potentially immense cost of systemic risk created by supply chain failures, there are several policy implications of this work. Considering the high level of turnover of supply relations in actual SCNs, it might be sufficient to introduce a tiny incentive toward lower-risk configurations that lead to significant systemic risk reduction. Such incentives could be similar to suggestions of topological risk mitigation in financial systems, such as systemic-risk insurance schemes  \cite{Poledna2016SRtax,leduc2017incentivize}. 
To make such schemes workable it is necessary to monitor systemic risk on the firm level. Except for a very few countries that are able to perform such monitoring, supply chain data at the firm level are not available and if they are, they are certainly not accessible to firms. To provide incentives for reasonable SCN rewiring in ways that market mechanisms naturally reach nationwide systemic risk optimal situations, it is necessary to think of ways of making this type of information available without negatively affecting consumer prices or the competitiveness of firms. Exploring implementable directions is subject to future work.

\section*{Materials and Methods} 
\subsection*{Economic Systemic Risk Index} To estimate the systemic risk content of a given production network we use the firm average Economic Systemic Risk Index (ESRI) \cite{Diem2022ESRI}. Each firm, $i$, in the network is assumed to produce according to its generalized Leontief production function,
\begin{equation}
    x_i= \min\left[
    \min_{k\in\mathcal{I}_i^{\text{es}}}\left[
        \frac{1}{\alpha_{ik}}\Pi_{ik}
        \right],
        \bar \beta_i + \frac{1}{\alpha_i}\sum_{k\in\mathcal{I}_i^{\text{ne}}}\Pi_{ik},
        \frac{1}{\alpha_{l_i}}l_i, \frac{1}{\alpha_{c_i}}c_i, 
    \right] .
\end{equation}
where $\Pi_{ik}=\sum_j W_{ji}\delta_{p_j,k}$ is the amount of input $k$ firm $i$ uses for production, $\mathcal{I}_i^{\text{es}}$ and $\mathcal{I}_i^{\text{ne}}$ represent the set of essential and non-essential inputs of firm $i$, and $l_i$ and $c_i$ are $i$'s labor and capital inputs. Parameter $\bar \beta_i$ is the production level possible without non-essential inputs, and $\alpha$ is the matrix of technological coefficients. These parameters are calibrated on the empirically observed inputs of every firm.
Every firm produces a single product described by its economic activity classification (NACE 3-digit). Whether a link constitutes an essential input or not is determined according to an expert-based survey for 56 sectors conducted by \cite{pichler2022IOmodel}. The survey identifies industries that provide inputs that other industries cannot easily substitute on short notice. The ESRI value of a firm is determined by setting the production level of the firm to zero and measuring the impact of this reduction both upstream and downstream of the firm. The resulting loss of total output is used to measure the systemic importance of the firm subject to the initial shock. ESRI$_i$, is the fraction of total production losses in the entire production network as a consequence of the failure of firm $i$. For more details, consult the SI in \cite{Diem2022ESRI}.

\subsection*{Supply chain networks} The six networks used in the analysis are derived from national VAT payments, representing the yearly aggregate taxes paid on firm-to-firm transactions in 2017 for Hungary and in 2015 for Ecuador. From these nationwide VAT graphs, we extract six subgraphs. Specifically, from the Ecuadorian network, we obtain two weighted graphs: one with 1,075 firms and 7,128 links (Crustaceans) and another with 890 firms and 9,323 links (Soft Drinks), along with their unweighted counterparts. Similarly, from the Hungarian network, we derive two unweighted graphs: one with 1,062 firms and 2,342 links (Food Production) and another with 1,147 firms and 2,561 links (Automotive). The subnetworks are extracted with the aim of analyzing specific production subsystems. We employ two methods, one based on identifying clusters of nodes using a greedy modularity optimization for the Hungarian network, and one based on a set of seed nodes for which Tier-1 suppliers and customers from relevant industries were then added in the case of Ecuador. These strategies have been designed to guarantee that the smaller subsystems have sufficient redundancy -- both in terms of firms with similar economic activities and in the links connecting them -- to ensure that internal reorganization is possible. We refer to the SI section~\ref{SI:extractECU} and~\ref{SI:extractHUN} for more details. 

\subsection*{Metropolis-Hastings algorithm and simulated annealing} 
A link rewiring step from a given network configuration, $a$, is proposed that would lead to a new configuration, $b$.
The decision to accept or reject the proposed rewiring step depends on whether it brings the network to a state characterized by lower systemic risk. After evaluating the average value of the $\text{ESRI}$ in $b$, the probability of accepting the rewiring step is
\begin{equation}
    p_{a\to b} = \min\{1, \exp(-\beta\,\Delta E_{ab})\} \quad , 
\end{equation}
where
\begin{equation}
    \Delta E_{ab} = \langle\text{ESRI}\rangle_b - \langle\text{ESRI}\rangle_a \quad . 
\end{equation}
$\beta=1/T$ is sometimes called an ``inverse temperature'', it is a parameter that controls the likelihood of accepting changes that increase the systemic risk, allowing the system to jump over local minima. We also employ a simulated annealing procedure, that often ameliorates the problem of getting stuck in local minima. To this end, we increase $\beta$ continuously with the number of iterations in the simulation.

\bibliography{bibliography}

\section*{Acknowledgements}
We are indebted to Tobias Reisch and Andras Borsos for making the Hungarian data available and to Pablo Astudillo-Estevez for the Ecuadorian data. The work was supported by the Austrian Science Foundation project ReMASS and the Austrian Federal Minister of Economy and Tourism via the Austrian Supply Chain Intelligence Institute. 

\section*{Author contributions statement}
ST conceived the work. GZ wrote the code and performed the data analysis. All authors developed the methodology, interpreted the results, and wrote the paper.

\vfill

\pagebreak
\onecolumngrid
\FloatBarrier

\section*{Supplementary Information}

\renewcommand{\thefigure}{S\arabic{figure}}
\setcounter{figure}{0}
\renewcommand{\thetable}{S\arabic{table}}
\setcounter{table}{0}
\renewcommand{\thesubsection}{S\arabic{subsection}}
\setcounter{subsection}{0} 
\renewcommand{\thesubsubsection}{\Alph{subsubsection}}
\setcounter{subsubsection}{0} 

\FloatBarrier
\subsection{Derivation of the six production networks}\label{sec:subNWs_derivation}

Our analysis is based on nationwide supply chain network (SCN) data from Ecuador and Hungary. These datasets are derived from value-added tax (VAT) payment records, capturing firm-to-firm transactions, with each direction of payment between firms (i.e., $i\to j$ and $j\to i$) recorded as a separate entry.". Firms are anonymized in both datasets, but their economic activity classifications are available - ISIC codes \cite{isicclassification2006} for Ecuador and NACE codes \cite{naceclassification2006} for Hungary. The transactions are aggregated annually, and data are available for multiple years. For Ecuador we have access to the transactions' aggregated volume, while for Hungary only the information about the existence of the transactions is known. Because the size of the nationwide SCNs is too large to simulate the disruption cascade of each node failure at every Monte Carlo rewiring step, we derive two weighted subnetworks from the Ecuadorian data (2015), their two unweighted counterparts, and two unweighted subnetworks from the Hungarian data (2017).

\subsubsection{Extracting two (plus two) production networks from the countrywide Ecuadorian SCN}\label{SI:extractECU}
The Ecuadorian dataset consists of VAT payments from 2015, capturing $\sim 14$ million yearly-aggregated transactions between $\sim 1$ million distinct anonymized taxpayers. Each row represents the total monetary flow from one taxpayer to another. To refine the data, we excluded transactions involving less than $3,000$ USD, self-loops (transactions where the payer and payee are the same), and entries involving entities labeled as \textit{not defined}, \textit{no economic activity}, or requiring \textit{verification}. Furthermore, we removed entries belonging to the \textit{personas naturales} type of tax contributor. After these filters, the dataset was reduced to almost $650$ thousand transactions between $\sim 65$ thousand taxpayers. The cleaned data were then used to construct a weighted directed SCN where the direction of the links is opposite to the flow of money, and the link weight is interpreted as the volume of goods exchanged. In this network, a firm's in-neighbors (out-neighbors) represent its suppliers (customers). The firms are originally categorized according to their economic activity in the ISIC scheme, which we converted to NACE.

The crustaceans network was derived from the 65k-firms network as follows. We notice that Ecuador's 2015 export basket \footnote{Ecuador's basket export in 2015: \href{https://atlas.cid.harvard.edu/explore?country=67&queryLevel=location&product=undefined&year=2015&productClass=HS&target=Product&partner=undefined&startYear=undefined}{atlas.cid.harvard.edu}} from the Atlas of Economic Complexity \cite{atlasEcoCompl} identifies ``Crustaceans'' and ``Prepared or preserved fish'' as contributing approximately 12.35\% of the country’s gross exports. This significant share suggests that the national production network is rich and developed enough to try to capture it through the extraction of a subgraph. First, within the 65k firms filtered SCN, we selected the 109 firms belonging to NACE 4-digit \textit{class} C1020 (``Processing and preserving of fish, crustaceans, and molluscs''). Second, we considered the set of all their in-neighbors (suppliers), excluding those also classified under C1020. The NACE 3-digit \textit{groups} of this set of firms were ranked based on their overrepresentation relative to their distribution in the 65-k firms SCN. From this ranking, we selected the top sixteen groups, ensuring that each contained at least five firms. This minimum threshold helped mitigate small-network effects, where having only a few representatives of a given NACE group in the subnetwork could artificially inflate their systemic importance due to their low replaceability within the subset. Third, we applied the same procedure to the out-neighbors (customers) of the C1020 firms, selecting the eight most overrepresented NACE groups. Fourth, we induced a subnetwork from the larger 65k-firms SCN, comprising the C1020 firms, their Tier 1 suppliers and customers from the selected groups, and all the links between these nodes. Finally, we considered the largest weakly connected component of this subnetwork. This approach, selecting overrepresented groups, was intentionally tailored to yield a network of manageable size while emphasizing economic activities that characterize the companies in the NACE class C1020 and distinguish this production network from the broader national supply chain network.

The derivation of the soft drinks network followed the same procedure, focusing on different economic activities. Here, we targeted the nodes in NACE 4-digit class C1107 (Manufacture of soft drinks; production of mineral waters and other bottled waters), whose output should represent a final product requiring no further processing. The NACE 3-digit groups of Tier 1 suppliers and customers for firms in C1107 exhibited greater diversity, reflecting the complexity expected for highly processed products. For the induced subgraph, we selected nodes from the top 23 overrepresented supplying NACE groups and the top 20 buying groups. Note that the links in these two subnetworks are weighted. Deriving the unweighted version of the two networks is trivial, we removed the information on the link weights simplifying their values to one.
\newpage
\subsubsection{Extracting two production networks from the countrywide Hungarian SCN}\label{SI:extractHUN}
The Hungarian dataset is also derived from payments between companies based in the country to the tax authorities, reporting supply invoices for VAT purposes. The identity of the firms is anonymized and not available to us. The data are available for every year, starting from 2015. Here we considered 2017, and the transactions are aggregated for the entire year. In 2017, only payments with tax content exceeding the threshold of 1 million HUF ($\sim3500$ USD, in that year) were required to be reported. In this work we do not have access to the link weights, and this means that the directed links in the resulting network capture only the existence of supply links, between companies exchanging goods or services with tax value above the threshold.
As in the Ecuadorian case, the size of the countrywide network is too large to evaluate $\langle \text{ESRI}\rangle_f - \langle \text{ESRI}\rangle_i$ at every Monte Carlo step, so we extract two subnetworks in the following way. We first note that firms in manufacturing (NACE C) are very heterogeneous in the inputs that are essential to them for production (Fig.9 in \cite{pichler2022IOmodel}). Therefore the subnetwork of manufacturing firms already displays many of the properties of the larger network in terms of the complexity of the dependency structure.
Hence, we isolate all the Hungarian firms in NACE section C and consider only the links between these firms. The induced subnetwork, is partitioned in communities with a Clauset-Newman-Moore greedy modularity maximization. After this step, we chose two communities with approximately 1000 nodes, and we looked at their composition in terms of NACE sectors. In Tab~\ref{tab:NACE3 nodes in the nws}, we break down the composition of the communities in terms of NACE 3-digit groups. The names for the two communities, ``food production'' network and ``automotive'' network, were chosen after the most overrepresented sectors in the subnetworks compared to the original countrywide NW.

\subsubsection{Properties of the networks}
The distributions of NACE economic activities for the derived production networks are presented in Table~\ref{tab:NACE3 nodes in the nws}. The rows break down the composition of each subnetwork, reporting the absolute number of firms categorized in each NACE group and the equivalent fraction of the network (in percentage). The most populated NACE groups are bolded for each network.

\begin{figure}[ht]
    \centering
    \vspace{1.5cm}
    \includegraphics[width=500pt]{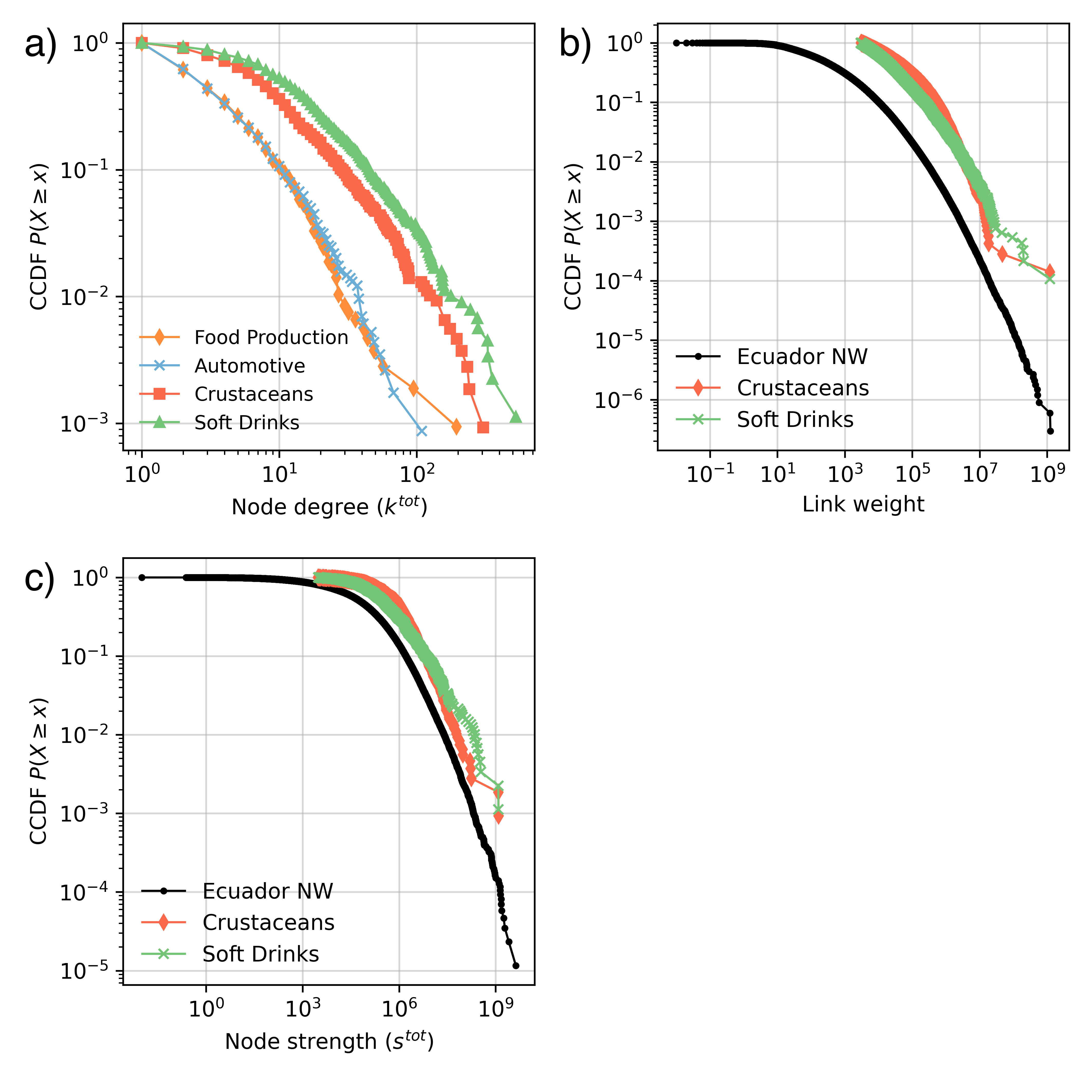}
    \caption{Complementary cumulative distribution functions (CCDFs) for several network properties. Panel \textbf{a)} reports the CCDFs of node total degree ($k^{tot}$) for the crustaceans, soft drinks, food production, and automotive networks. Node degrees do not change between the weighted and unweighted versions of the networks. In panel \textbf{b)}, CCDF of node total strength ($s^{tot}$) comparing the Ecuadorian weighted subnetworks (crustaceans, soft drinks), and the original nationwide one (Ecuador network). In panel \textbf{c)}, CCDF of link weight for the same weighted networks.}
    \label{fig:ccdfs}
\end{figure}

In Fig.~\ref{fig:ccdfs}, we reported the Complementary Cumulative Distribution Function (CCDF) for several node and link attributes on a log-log scale. In Fig.~\ref{fig:ccdfs} \textbf{a)}, we plotted the CCDF of the node total degrees $k^{tot}$. The different colored lines correspond to the different networks. A clear pattern is visible, with the crustaceans and soft drinks lines positioned further to the right compared to the food production and automotive lines. This shift reflects the fact that nodes in these networks tend to have higher degrees. The maximum total degrees observed in the food production, automotive, crustaceans, and soft drinks networks are 195, 109, 305, and 527, respectively. By construction, the rewiring algorithm in the case of unweighted networks preserves in-degrees, out-degrees, and total degrees, so the plot does not change after the risk mitigation.
Fig.~\ref{fig:ccdfs} \textbf{b)} illustrates the CCDF of link weights for the weighted Ecuadorian subnetworks, and compares them with the CCDF in the original nationwide SCN (Ecuador NW in the legend). As shown by the maximum x-axis (weight) values of the three lines, both induced subgraphs include links from the tail of the main distribution, overlapping with the higher-weight region of the Ecuador NW curve. The threshold on link weights applied before the subnetwork derivation introduces a truncation effect along the x-axis, preventing links with weight lower than $3,000$ USD from appearing in the induced graphs. Fig.~\ref{fig:ccdfs} \textbf{c)} presents the CCDF of node total strength $s^{tot}$ for the same weighted networks. Here, the truncation on link weights affects node strengths reducing their overall values and preventing the existence of nodes with strengths below $3,000$.

\newpage
\begin{longtable}{lrrrr} 
    \toprule
     & Food production & Automotive& Crustaceans & Soft drinks \\
    NACE 3-digit groups &  &  &  &  \\
    \midrule
    031 Fishing & - & - & \textbf{98 (9.1\%)} & 8 (0.9\%) \\
    032 Aquaculture & - & - & \textbf{422 (39.3\%)} & - \\
    101 Processing and preserving of meat and production o & \textbf{68 (6.4\%)} & - & - & - \\
    102 Processing and preserving of fish, crustaceans and & - & - & \textbf{97 (9.0\%)} & 13 (1.5\%) \\
    103 Processing and preserving of fruit and vegetables & 41 (3.9\%) & - & - & 6 (0.7\%) \\
    104 Manuf. of vegetable and animal oils and fats & 4 (0.4\%) & - & - & - \\
    105 Manuf. of dairy products & 25 (2.4\%) & - & - & 9 (1.0\%) \\
    106 Manuf. of grain mill products, starches and starch & 22 (2.1\%) & - & - & - \\
    107 Manuf. of bakery and farinaceous products & \textbf{103 (9.7\%)} & 1 (0.1\%) & - & 7 (0.8\%) \\
    108 Manuf. of other food products & 47 (4.4\%) & - & - & 22 (2.5\%) \\
    109 Manuf. of prepared animal feeds & 48 (4.5\%) & - & 12 (1.1\%) & - \\
    110 Manuf. of beverages & \textbf{102 (9.6\%)} & 3 (0.3\%) & 29 (2.7\%) & \textbf{63 (7.1\%)} \\
    120 Manuf. of tobacco products & 1 (0.1\%) & - & - & - \\
    133 Finishing of textiles & 1 (0.1\%) & - & - & - \\
    139 Manuf. of other textiles & 8 (0.8\%) & 8 (0.7\%) & - & - \\
    141 Manuf. of wearing apparel, except fur apparel & 5 (0.5\%) & 4 (0.3\%) & - & - \\
    142 Manuf. of articles of fur & 1 (0.1\%) & - & - & - \\
    143 Manuf. of knitted and crocheted apparel & 1 (0.1\%) & - & - & - \\
    152 Manuf. of footwear & - & 1 (0.1\%) & - & - \\
    161 Sawmilling and planing of wood & 15 (1.4\%) & 12 (1.0\%) & - & - \\
    162 Manuf. of products of wood, cork, straw and plaiti & 28 (2.6\%) & 12 (1.0\%) & - & - \\
    171 Manuf. of pulp, paper and paperboard & 5 (0.5\%) & - & 7 (0.7\%) & - \\
    172 Manuf. of articles of paper and paperboard  & \textbf{62 (5.8\%)} & 16 (1.4\%) & 10 (0.9\%) & 10 (1.1\%) \\
    181 Printing and service activities related to printin & 51 (4.8\%) & 15 (1.3\%) & - & 22 (2.5\%) \\
    182 Reproduction of recorded media & 1 (0.1\%) & 1 (0.1\%) & - & - \\
    192 Manuf. of refined petroleum products & 2 (0.2\%) & - & - & - \\
    201 Manuf. of basic chemicals, fertilisers and nitroge & 15 (1.4\%) & 2 (0.2\%) & - & 31 (3.5\%) \\
    202 Manuf. of pesticides and other agrochemical produc & 5 (0.5\%) & - & - & - \\
    203 Manuf. of paints, varnishes and similar coatings,  & 11 (1.0\%) & 2 (0.2\%) & - & - \\
    204 Manuf. of soap and detergents, cleaning and polish & 11 (1.0\%) & - & - & - \\
    205 Manuf. of other chemical products & 14 (1.3\%) & 2 (0.2\%) & 22 (2.0\%) & 16 (1.8\%) \\
    206 Manuf. of man-made fibres & 1 (0.1\%) & - & - & - \\
    212 Manuf. of pharmaceutical preparations & - & 2 (0.2\%) & - & - \\
    221 Manuf. of rubber products & 2 (0.2\%) & 26 (2.3\%) & - & - \\
    222 Manuf. of plastic products & \textbf{136 (12.8\%)} & \textbf{77 (6.7\%)} & 28 (2.6\%) & 44 (4.9\%) \\
    231 Manuf. of glass and glass products & 1 (0.1\%) & 1 (0.1\%) & - & - \\
    232 Manuf. of refractory products & - & 2 (0.2\%) & - & - \\
    233 Manuf. of clay building materials & 2 (0.2\%) & - & - & - \\
    234 Manuf. of other porcelain and ceramic products & - & 1 (0.1\%) & - & - \\
    235 Manuf. of cement, lime and plaster & 2 (0.2\%) & - & - & - \\
    236 Manuf. of articles of concrete, cement and plaster & 14 (1.3\%) & 2 (0.2\%) & - & - \\
    237 Cutting, shaping and finishing of stone & 2 (0.2\%) & - & - & - \\
    239 Manuf. of abrasive products and non-metallic miner & 8 (0.8\%) & 1 (0.1\%) & - & - \\
    241 Manuf. of basic iron and steel and of ferro-alloys & - & 1 (0.1\%) & - & - \\
    243 Manuf. of other products of first processing of st & 1 (0.1\%) & 2 (0.2\%) & - & - \\
    244 Manuf. of basic precious and other non-ferrous met & - & 7 (0.6\%) & - & - \\
    245 Casting of metals & - & 29 (2.5\%) & - & - \\
    251 Manuf. of structural metal products & 23 (2.2\%) & \textbf{76 (6.6\%)} & - & - \\
    252 Manuf. of tanks, reservoirs and containers of meta & 2 (0.2\%) & 4 (0.3\%) & - & - \\
    253 Manuf. of steam generators, except central heating & 1 (0.1\%) & - & - & - \\
    255 Forging, pressing, stamping and roll-forming of me & - & 17 (1.5\%) & - & - \\
    256 Treatment and coating of metals; machining & 26 (2.4\%) & \textbf{247 (21.5\%)} & - & - \\
    257 Manuf. of cutlery, tools and general hardware & 3 (0.3\%) & \textbf{76 (6.6\%)} & - & - \\
    259 Manuf. of other fabricated metal products & 10 (0.9\%) & 27 (2.4\%) & - & - \\
    261 Manuf. of electronic components and boards & 3 (0.3\%) & 27 (2.4\%) & - & - \\
    262 Manuf. of computers and peripheral equipment & - & 2 (0.2\%) & - & - \\
    263 Manuf. of communication equipment & 3 (0.3\%) & 7 (0.6\%) & - & - \\
    265 Manuf. of instruments and appliances for measuring & 6 (0.6\%) & 17 (1.5\%) & 5 (0.5\%) & - \\
    266 Manuf. of irradiation, electromedical and electrot & - & 1 (0.1\%) & - & - \\
    267 Manuf. of optical instruments and photographic equ & - & 2 (0.2\%) & - & - \\
    271 Manuf. of electric motors, generators, transformer & 3 (0.3\%) & 11 (1.0\%) & - & - \\
    273 Manuf. of wiring and wiring devices & 3 (0.3\%) & 9 (0.8\%) & - & - \\
    274 Manuf. of electric lighting equipment & 2 (0.2\%) & 6 (0.5\%) & - & - \\
    275 Manuf. of domestic appliances & - & 3 (0.3\%) & - & - \\
    279 Manuf. of other electrical equipment & 2 (0.2\%) & 9 (0.8\%) & - & - \\
    281 Manuf. of general-purpose machinery & 8 (0.8\%) & 24 (2.1\%) & 7 (0.7\%) & - \\
    282 Manuf. of other general-purpose machinery & 22 (2.1\%) & 55 (4.8\%) & - & 16 (1.8\%) \\
    283 Manuf. of agricultural and forestry machinery & 2 (0.2\%) & 5 (0.4\%) & - & - \\
    284 Manuf. of metal forming machinery and machine tool & - & 20 (1.7\%) & - & - \\
    289 Manuf. of other special-purpose machinery & 21 (2.0\%) & 50 (4.4\%) & 6 (0.6\%) & - \\
    291 Manuf. of motor vehicles & 1 (0.1\%) & 5 (0.4\%) & - & - \\
    292 Manuf. of bodies (coachwork) for motor vehicles; m & - & 3 (0.3\%) & - & - \\
    293 Manuf. of parts and accessories for motor vehicles & - & \textbf{77 (6.7\%)} & - & - \\
    301 Building of ships and boats & - & - & 12 (1.1\%) & - \\
    302 Manuf. of railway locomotives and rolling stock & 2 (0.2\%) & 4 (0.3\%) & - & - \\
    303 Manuf. of air and spacecraft and related machinery & - & 1 (0.1\%) & - & - \\
    309 Manuf. of transport equipment n.e.c. & 1 (0.1\%) & - & - & - \\
    310 Manuf. of furniture & 1 (0.1\%) & 43 (3.7\%) & - & - \\
    323 Manuf. of sports goods & - & 1 (0.1\%) & - & - \\
    324 Manuf. of games and toys & 1 (0.1\%) & 2 (0.2\%) & - & - \\
    325 Manuf. of medical and dental instruments and suppl & 1 (0.1\%) & 5 (0.4\%) & - & - \\
    329 Manufacturing n.e.c. & 5 (0.5\%) & 4 (0.3\%) & - & - \\
    331 Repair of fabricated metal products, machinery and & 31 (2.9\%) & 52 (4.5\%) & 20 (1.9\%) & - \\
    332 Installation of industrial machinery and equipment & 13 (1.2\%) & 25 (2.2\%) & - & - \\
    351 Electric power generation, transmission and distri & - & - & - & 10 (1.1\%) \\
    360 Water collection, treatment and supply & - & - & - & 5 (0.6\%) \\
    370 Sewerage & - & - & - & 6 (0.7\%) \\
    451 Sale of motor vehicles & - & - & - & 20 (2.2\%) \\
    463 Wholesale of food, beverages and tobacco & - & - & \textbf{84 (7.8\%)} & \textbf{73 (8.2\%)} \\
    466 Wholesale of other machinery, equipment and suppli & - & - & - & \textbf{71 (8.0\%)} \\
    469 Non-specialised wholesale trade & - & - & - & 42 (4.7\%) \\
    471 Retail sale in non-specialised stores & - & - & - & \textbf{51 (5.7\%)} \\
    472 Retail sale of food, beverages and tobacco in spec & - & - & 9 (0.8\%) & 13 (1.5\%) \\
    473 Retail sale of automotive fuel in specialised stor & - & - & - & 26 (2.9\%) \\
    477 Retail sale of other goods in specialised stores & - & - & - & 14 (1.6\%) \\
    502 Sea and coastal freight water transport & - & - & 21 (2.0\%) & - \\
    512 Freight air transport and space transport & - & - & 9 (0.8\%) & - \\
    521 Warehousing and storage & - & - & - & 7 (0.8\%) \\
    522 Support activities for transportation & - & - & \textbf{149 (13.9\%)} & - \\
    552 Holiday and other short-stay accommodation & - & - & - & 47 (5.3\%) \\
    561 Restaurants and mobile food service activities & - & - & - & \textbf{73 (8.2\%)} \\
    562 Event catering and other food service activities & - & - & 10 (0.9\%) & 12 (1.3\%) \\
    563 Beverage serving activities & - & - & 11 (1.0\%) & 13 (1.5\%) \\
    601 Radio broadcasting & - & - & - & 7 (0.8\%) \\
    651 Insurance & - & - & - & 12 (1.3\%) \\
    712 Technical testing and analysis & - & - & 7 (0.7\%) & - \\
    731 Advertising & - & - & - & 46 (5.2\%) \\
    732 Market research and public opinion polling & - & - & - & 6 (0.7\%) \\
    781 Activities of employment placement agencies & - & - & - & 5 (0.6\%) \\
    852 Primary education & - & - & - & 9 (1.0\%) \\
    854 Higher education & - & - & - & 5 (0.6\%) \\
    931 Sports activities & - & - & - & 50 (5.6\%) \\
    \bottomrule
    \\
    \caption{
    Cross-sectoral distribution of firms by NACE 3-digit groups across the networks (food production, automotive, crustaceans, and soft drinks). Percentages in parentheses represent the share of nodes within the given NACE category. For each network, bolded values highlight the most populated categories.
    }
    \label{tab:NACE3 nodes in the nws}
\end{longtable}
\FloatBarrier
\newpage

\FloatBarrier

\pagebreak
\subsection{Economic Systemic Risk Index (ESRI)}\label{SI:sectESRI}
The Economic Systemic Risk Index (ESRI) quantifies a firm's systemic importance within a supply chain network by measuring the potential cascade of production failures triggered when the firm suddenly ceases operations.
In the networks, each firm is represented as a node that produces goods or services (``products'') using the entirety of the inputs sourced from suppliers (in-neighbors) and selling outputs to customers (out-neighbors). In- and out-neighbors are determined by the direction of the links, which follow the flow of products, opposite to the direction of the payments in the dataset.
Since product-level information is unavailable, we use the NACE 3-digit classification of each firm as a proxy for the products it produces. This implies that firms within the same NACE group can supply the same products.
The weight of the link from supplier $j$ to customer $i$, denoted $W_{ji}$, represents the monetary value of the supplied product, serving as an estimate of the amount of product.
Every firm is equipped with a generalized Leontief production function (GLPF), that models how the intermediate input quantities are transformed into output while distinguishing between essential and non-essential inputs. We employed the results of the survey conducted for \cite{pichler2022IOmodel} that classifies which NACE 2-digit division inputs are essential, non-essential, or irrelevant for production in each division. The possible production in each firm given input volumes and types is:
\begin{equation}
    x_i= \min\left[
    \min_{k\in\mathcal{I}_i^{\text{es}}}\left[
        \frac{1}{\alpha_{ik}}\Pi_{ik}
        \right],
        \bar \beta_i + \frac{1}{\alpha_i}\sum_{k\in\mathcal{I}_i^{\text{ne}}}\Pi_{ik},
        \frac{1}{\alpha_{l_i}}l_i, \frac{1}{\alpha_{c_i}}c_i 
    \right] .
\label{eq:GL prod func}
\end{equation}
where $\Pi_{ik}=\sum_j W_{ji}\delta_{p_j,k}$ is the amount of input $k$ firm $i$ uses for production, $\mathcal{I}_i^{\text{es}}$ and $\mathcal{I}_i^{\text{ne}}$ represent the set of essential and non-essential inputs of firm $i$, and $l_i$ and $c_i$ are $i$'s labor and capital inputs, which are untouched and ignored in our work settings. Essential inputs are treated in a Leontif way, and non-essential inputs affect production linearly. Parameter $\bar \beta_i$ is the production level possible without non-essential inputs, and $\alpha$ is the matrix of technological coefficients. This formulation captures firms' heterogeneity in input usage and substitutability.

To calculate ESRI of a firm $j$, we initialize the network by calibrating production functions at time $t=0$. The production output that each firm $i$ is able to sustain with the inputs delivered from the suppliers (calculated with \eqref{eq:GL prod func}) is exactly equal to the summed volume of its sales transactions, the node outstrength $s_i^{out}$. Every firm $i$ operates at full capacity, measured as $h_i(t=0)=1$. At $t=1$, the exogenous shock hits and firm $j$ ceases operations. This triggers both an upstream shock to suppliers and a downstream shock to customers. Firm $j$'s suppliers face reduced demand and adjust their production level $h(t=1)$, decreasing their output by the fraction that was supplied to $j$. Firm $j$'s customers decrease production depending on the essentiality of the missing input in their production function. Essential inputs propagate shocks in a non-linear (Leontief) way, while non-essential inputs do so linearly. The impact of the lack of input in the production function is softened depending on how easy is to replace firm $j$ as a supplier. The market share is used as a proxy for this replaceability: firms with higher market share are assumed to be harder to replace. 

The shock propagates iteratively upstream and downstream, distributing the shock proportionally between the neighbors until convergence at time $T$. At $T$, the firms have decreased their empirical production level $s^{out}$ by a factor $1-h(T)$, and the ESRI of firm $i$ is defined as the total production loss in the network:
\begin{equation}
    \text{ESRI}_j= \sum_{i=1}^n \frac{s_i^{out}}{\sum_{l=1}^n s_l^{out}} \left( 1- h_i(T)\right) \quad .
\label{eq:ESRIdef}
\end{equation}
Since rewiring is not permitted during cascade propagation, ESRI does not reflect dynamic adaptation over time. However, it is not designed to simulate real-world shock propagation in full detail, but rather to measure and provide insights into systemic relevance and the network’s potential susceptibility to firm-level disruptions.

\newpage

\subsection{Details of the rewiring algorithm}\label{SI:sectREWIRINGRULE}
\subsubsection{Firms' production constraints require considering multiple links}
We developed a Monte Carlo link-swapping algorithm to rewire weighted and unweighted networks while maintaining specific constraints. This method ensures that some supply chain networks' properties and firms' production capabilities remain consistent throughout the exploration of alternative configurations. In this paragraph, we provide a detailed explanation of the algorithm and its constraints, along with the extensions required to accommodate weighted networks.

Rewiring the network does not aim to model the temporal evolution of the firms forming it, but rather explore alternative configurations. For this reason, when rewiring a link, we aim to replace a supplier (customer) with another \textit{equivalent} firm providing (demanding) the same specific product or service. In practice, however, detailed product-level information is often unavailable. Instead, firms are usually classified according to broad industrial schemes such as NACE or ISIC using different levels of granularity. In this case, we use the economic activity classification of each firm pair forming an edge as a proxy for the specific good being exchanged. The level of aggregation we chose is the NACE 3-digit group. Thus, we assume that firms in the same NACE group supplying from firms in another specific NACE group can freely choose to change suppliers from this second set. Of course, the algorithm is perfectly suited to work with more detailed product information. Rewiring should not alter the technology and productivity of the firms and then it must preserve the observed ratios and volumes of intermediate goods used to produce the levels of output. Because of the strong correlation between a node's connectivity and its systemic importance, we should also not alter the number of suppliers and customers of a firm. Finally, if a firm is empirically observed to purchase from a given number of suppliers providing the same intermediate good, we do not want to change this level of supply redundancy, which is a property of the firm. Similarly, the set of sectors a firm sells to and the number of customers in each sector should remain unchanged.

From a network perspective, these constraints amount to preserving both the in and out degree and strength per sector of each node. For simplicity consider first an unweighted network where each link has the same value of one unit. In this case, the degree and strength constraints are equivalent. These rules require considering more than one link in each rewiring step because deleting, introducing, or cutting and redirecting one single link would inevitably change the number of connections of its two old or new ends. The problem can be solved by rewiring multiple links rather than rewiring them individually. For simplicity, we considered two links per rewiring event. After selecting two links representing the same product - in our case with the same combination of (source-sector, target-sector), the suppliers of the two target nodes are exchanged. The same swap has a complementary point of view, in which the customers of the two source nodes are exchanged. For the illustration of the swap-rewiring, refer to Fig.2 of the main text. The swap is reversible, and this is important in the exploration of the configuration space as it guarantees that we can explore all possible network states without getting stuck.

This simple swap-rewiring has limitations: while degree and strength are strongly correlated, there can be significant heterogeneity of link weights. As such the simple swap discussed here does not preserve the production level of firms as strictly as we would like. However, we can alter the algorithm in such a way as to be more flexible on the degrees in exchange for tighter bounds on the strengths. To do so, we keep the same algorithm for the initial proposal of the swap. If the weights of two selected links differ, we look at the size of their difference: if the difference is sufficiently small, we deem the edges equivalent and perform the swap as usual. We considered sufficiently small a value below the threshold for the links in the original Ecuadorian supply chain network, 3000 USD. Note that to preserve the total weight of the graph, when performing the swap we will ensure that the in-strengths of the nodes are exactly kept, while the out-strengths can change slightly. This can be seen as anchoring the weight of the edge from the customer side and not from the supplier side. In production words, this means that the input levels of firms cannot change, while the output can.
To ensure that the algorithm does not significantly change the output coefficients and production level by small increments or reductions, the node out-strengths are allowed to deviate by no more than 20\% from their empirically observed values to accommodate this swap. If the move would result in a violation of this strength constraint it is rejected. The other case to consider is if the difference between the two weights is too large to just perform a simple swap. In this case, we must choose between not performing the swap at all or violating the degree constraint by introducing a new link. We choose the latter, as this ensures more configuration can be explored given the small number of links that are exactly of the same value. Note that also these swap moves are all reversible and as such we can freely explore if the addition or subtraction of this link increases or decreases the systemic risk level of the network. 

\begin{figure}[t]
    \centering
    \includegraphics[width=\textwidth]{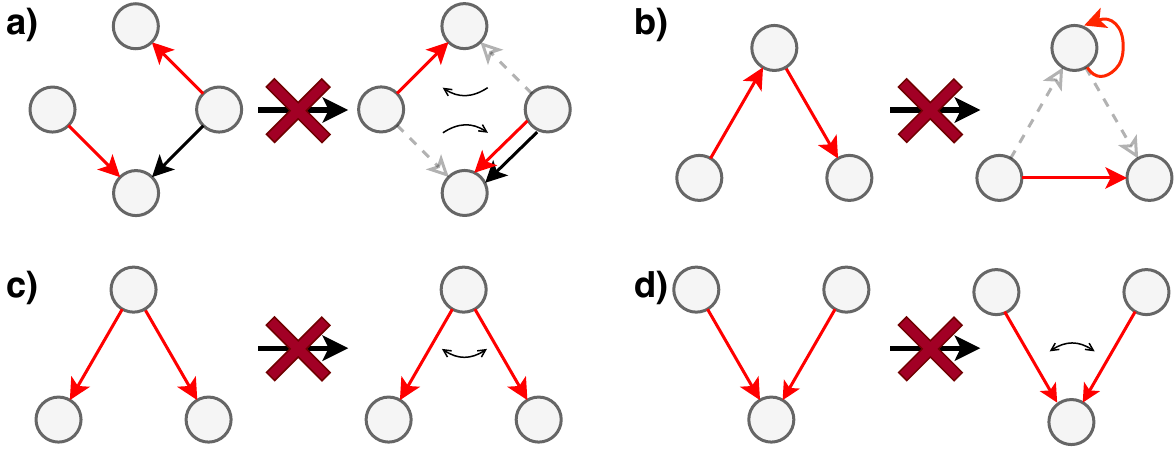}
    \caption{Illustration of the pathological cases we want to avoid in the swap-rewiring algorithm. In panel \textbf{a)}, we illustrate the potential introduction of a multiedge. The source in the first link should not be an in-neighbor of the target in the second one, or equivalently, the target in the first link should not be an out-neighbor of the source in the second one. Panel \textbf{b)} shows the scenario that would introduce self-loops. Panel \textbf{c)} and \textbf{d)} represent two trivial swaps that do not change the topology in unweighted networks.}
    \label{fig:pathological_cases}
\end{figure}
\FloatBarrier

\subsubsection{Sampling two links in the network.}
In each Monte Carlo step, two links in the network are rewired according to the scheme in Fig.2 in the main paper. For the case of unweighted networks, the two links are chosen in the following way. At the beginning of each rewiring event, the current network configuration is characterized by the set of nodes $\mathcal{N}$, that never changes, and the set of links $\mathcal{L}$. A first link $l_1$ is selected at random (with uniform probability), connecting $l_1^{source} \to l_1^{target}$. The second link $l_2$ should have the same combination of NACE 3-digit groups to avoid affecting the production functions of the nodes. This means that the source $l_2^{source}$ must belong to the set of nodes with the same NACE3 code of $l_1^{source}$, i.e., $\mathcal{S} = \{n\in\mathcal{N} \ | \ \text{NACE3}(n) = \text{NACE3}(l_1^{source})\}\subset\mathcal{N}$. From this set, we further exclude:
\begin{itemize}
    \item Node $l_1^{target}$, because otherwise we would introduce self-loops (see Fig.~\ref{fig:pathological_cases} \textbf{b)}). The set is restricted to: \[ \mathcal{S'} = \mathcal{S}\setminus \{l_1^{target}\}\]    
    \item For networks with binary links, in-neighbors of $l_1^{target}$ are also excluded from the set of possible sources. Choosing node $l_1^{source}$ again as the source in $l_2$ results in the trivial link swap between customers of the same firm in panel \textbf{c)}. Choosing another in-neighbor and proceeding with the swap would introduce a link that already exists, resulting in increasing its weight by 1 (panel \textbf{a)}). For unweighted networks, the set of possible sources is then: \[ \mathcal{S''} = \mathcal{S'}\setminus\{l_1^{source}\}\setminus\{n\to l_1^{target}\}\]
\end{itemize}
Similarly, the target in the second link should belong to $\mathcal{T} = \{n\in\mathcal{N} \ | \ \text{NACE3}(n) = \text{NACE3}(l_1^{target})\}$. Here, we exclude:
\begin{itemize}
    \item node $l_1^{source}$, otherwise the introduction of self-loops (panel \textbf{b)}): \[ \mathcal{T'} = \mathcal{T}\setminus\{l_1^{source}\}\]
    \item For networks with binary links, out-neighbors of $l_1^{source}$ are ruled out. Node $l_1^{target}$ leads to ineffective link-swap between suppliers (panel \textbf{d)}), and choosing other out-neighbors of $l_1^{source}$ increases link weights (panel \textbf{a)}). Finally, the target in $l_2$ should be picked from: \[\mathcal{T''} = \mathcal{T'}\setminus\{l_1^{target}\}\setminus\{l_1^{source}\to n\}\] 
\end{itemize}
Every link pointing from a node in $\mathcal{S''}$ to a node in $\mathcal{T''}$ is eligible as a second link for the swap, so we intersect the links going out from nodes in $\mathcal{S''}$ with the links ending to nodes in $\mathcal{T''}$ and we choose one of them with uniform probability. In case the set is empty, the algorithm goes back to the first step and samples again the first link $l_1$.

In the case of weighted networks, the considerations that brought to the sets $\mathcal{S'}$ and $\mathcal{T'}$ are still valid, but there is no reason to forbid increasing the weight of a link as in panel \textbf{a)} and restricting to $\mathcal{S'}$ and $\mathcal{T''}$. Moreover, the swap of volumes between customers and suppliers becomes meaningful, as the node out-strengths are allowed to change if they stay between 80\% and 120\% of the empirical value. The second link is chosen by sampling the set of links from nodes in $\mathcal{S'}$ to nodes in $\mathcal{T'}$.

\subsubsection{Pseudocode}
We schematize the rewiring algorithm in the following pseudocode. In the unweighted rewiring, we want to avoid the introduction of multi-edges (increasing the weight of the links) and to save computational time, we also exclude the trivial swap of customers (suppliers) in the same firm. Lines 7 and 9 in the {\scriptsize FIND2LINKs} prevent these scenarios, depicted in Fig.~\ref{fig:pathological_cases} \textbf{a)}, \textbf{c)} and \textbf{d)}.
After finding two links, procedure {\scriptsize REWIRING} handles the weighted-unweighted cases. The functions {\scriptsize SWAPPARTIAL} and {\scriptsize SWAPFULL} contain the operations on the two links discussed in detail in the first paragraph of this section.

\begin{figure}[h!]
    \centering
    \begin{algorithmic}[1]
    \Procedure{find2links}{$G$} 
    \While{\texttt{link2 = NULL}}
        \State \texttt{link1 $\gets$ sample(edges($G$))}
        \State \texttt{source, target $\gets$ ends(link1)}
        \State \texttt{SourceCandidates $\gets$ sameNACE3($G$, source) $\setminus$ target}
        \State \texttt{SourceCandidates $\gets$ SourceCandidates $\setminus$ inneighbors(target)} \Comment{In case of unweighted networks}
        
        \State \texttt{TargetCandidates $\gets$ sameNACE3($G$, target) $\setminus$ source}
        \State \texttt{TargetCandidates $\gets$ TargetCandidates $\setminus$ outneighbors(source)} \Comment{In case of unweighted networks}
        \State \texttt{OutgoingLinks $\gets$ edges($G$).from(CandidateSources)}
        \State \texttt{IncomingLinks $\gets$ edges($G$).to(CandidateTargets)}
        \State \texttt{bridges $\gets$ OutgoingLinks $\cap$ IncomingLinks}
        \State \texttt{link2 $\gets$ sample(bridges)}
    \EndWhile
    \State \textbf{return} \texttt{link1, link2}
    \EndProcedure
    \State
    \Procedure{rewiring}{$G$} 
    \While{\texttt{TRUE}}   
       \State \texttt{link1, link2} $\gets$ {\scriptsize FIND$2$LINKS}($G$)
       \State \texttt{residue $\gets$ |weight(link1) $-$ weight(link2)|}
       \If{\texttt{residue $>$ threshold}}
        \State \texttt{SwappedG $\gets$} {\scriptsize SWAPPARTIAL}\texttt{($G$, link1, link2)}
        \State \textbf{\texttt{break}}
       \ElsIf{{\scriptsize SUPPLIERS\_OUTSTRENGTHS\_CONSTRAINT}\texttt{(link1,link2)}}
       \State \texttt{SwappedG $\gets$} {\scriptsize SWAPFULL}\texttt{($G$, link1, link2)}
       \State \textbf{break}
       \EndIf
    \EndWhile
    \State \textbf{return} \texttt{SwappedG}
    \EndProcedure
    \end{algorithmic}
    \caption{Procedure {\scriptsize FIND$2$LINKS} in the pseudocode summarizes the searching algorithm for choosing two links in the weighted or unweighted network $G$. 
    Procedure {\scriptsize REWIRING} summarizes how to handle the cases for weighted or unweighted swap-rewiring.}\label{alg:}
\end{figure}
\FloatBarrier
\newpage

\subsection{Results for all the fixed $\beta$ and simulated annealing Monte Carlo simulations}\label{SI:sectALLBETA}
\begin{figure}[!h]
    \centering
    \includegraphics[width=450pt]{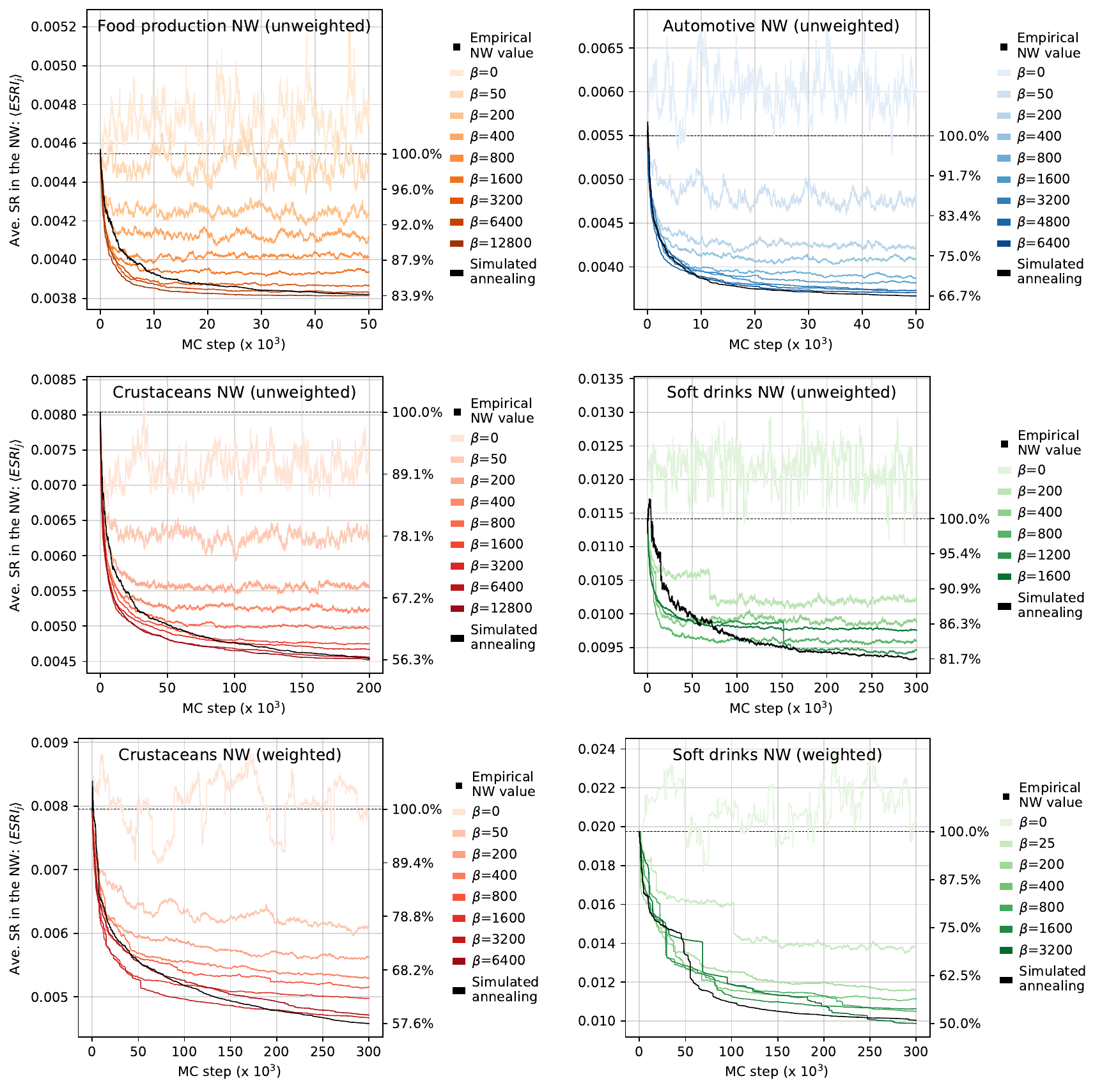}
    \caption{Evolution of $\langle\text{ESRI}\rangle$ along the Monte Carlo simulations for all the networks. The first two rows display results for unweighted networks, while the last row shows those for weighted networks. Each subfigure includes a trajectory from simulated annealing and one trajectory for each fixed $\beta$ value tested. Lowering the temperature ($T = 1/\beta$) generally decreases the convergence value, although excessively low $T$ can cause the exploration to get stuck. The fixed-$\beta$ results were used to calibrate the parameters for the simulated annealing procedures of Table~\ref{tab:simulated annealing procedures}.}
    \label{fig:all simulations}
\end{figure}

In the main text, we reported the results of three runs of the $\langle\text{ESRI}\rangle$ minimization at fixed $\beta$ and one simulated annealing run, for one weighted network (soft drinks) and one unweighted network (food production). Here, we attach the results for all the tested values of fixed $\beta$, for all the networks. In Fig.~\ref{fig:all simulations}, it is evident how the convergence value of $\langle\text{ESRI}\rangle$ depends on the $\beta=1/T$ temperature parameter. In each subfigure, only one trajectory per $\beta$ is reported. The simulated annealing temperature curve was calibrated after the results with fixed beta, and the chosen curves are reported in Table~\ref{tab:simulated annealing procedures}. Non trivially, when comparing $\langle\text{ESRI}\rangle$ values from the empirical data, those obtained from random explorations of the configuration space ($\beta = 0$), and the minimized states accessed through the Metropolis-Hastings algorithm, empirical networks are consistently closer to undriven random configurations than to the risk-mitigated states. This suggests that none of the analyzed networks can naturally reach the minimal $\langle\text{ESRI}\rangle$ states without implementing policies that promote some rewiring behavior.

\FloatBarrier

\begin{table}[h!]
    \centering
    \resizebox{\textwidth}{!}{
    \begin{tabular}{|c|cccccc|}
        \toprule
        Network & \thead{Food production\\(unweighted)} & \thead{Automotive\\(unweighted)} & \thead{Crustaceans\\(unweighted)} & \thead{Soft drinks\\(unweighted)} & \thead{Crustaceans\\(weighted)} & \thead{Soft drinks \\(weighted)}\\
        \midrule
        \thead{Simulated annealing\\$\beta$ curve} & $\beta(\text{step})=\frac{12800\cdot\text{step}}{50000}$ & $\beta(\text{step})=\frac{12800\cdot\text{step}}{50000}$ & $\beta(\text{step})=\frac{6400\cdot\text{step}}{200000}$ & $\beta(\text{step})=\frac{2100\cdot\text{step}}{300000}$ & $\beta(\text{step})=\frac{7200\cdot\text{step}}{300000}$ & $\beta(\text{step})=\frac{3600\cdot\text{step}}{300000}$ \\
        \bottomrule
    \end{tabular}
    }
    \caption{Temperature decrease curves used for the simulated annealing results.}
    \label{tab:simulated annealing procedures}
\end{table}
\FloatBarrier

\subsection{Comparison of the ESRI profiles in the best $\langle\text{ESRI}\rangle$ minimization results and relative empirical networks.}\label{SI:sectBESTPROFILES}
\begin{figure}[!h]
    \centering
    \includegraphics[width=400pt]{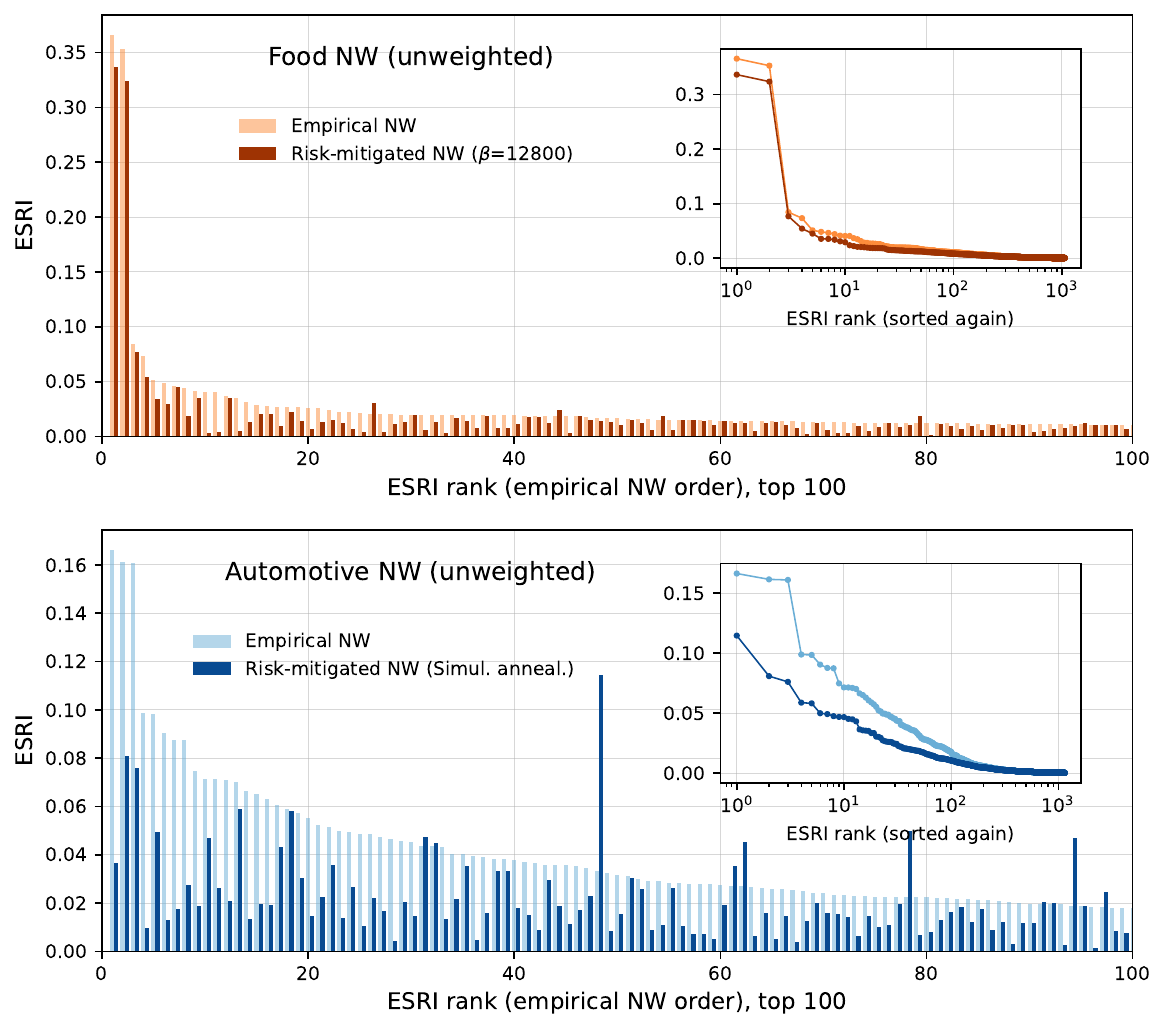}
    \caption{Systemic risk profiles of the food production network (top panel) and automotive network (bottom panel). The networks are considered in their unweigthed version. The light bars report firms' ESRI values in the empirical network, the darker bars the new ESRI values in the rewired networks. We consider the best risk-mitigation results with the largest reduction of $\langle \text{ESRI} \rangle$, fixed $\beta=12800$ and simulated annealing. Firms are ranked by their ESRI in the empirical network along the $x$-axis, displaying only the top 100 riskiest firms. The inset presents the same profiles, with the rewired profile now rank-ordered for direct comparison.}
    \label{fig:ESRI profiles bef/aft 1}
\end{figure}

\begin{figure}
    \centering
    \vspace{5cm}
    \includegraphics[width=400pt]{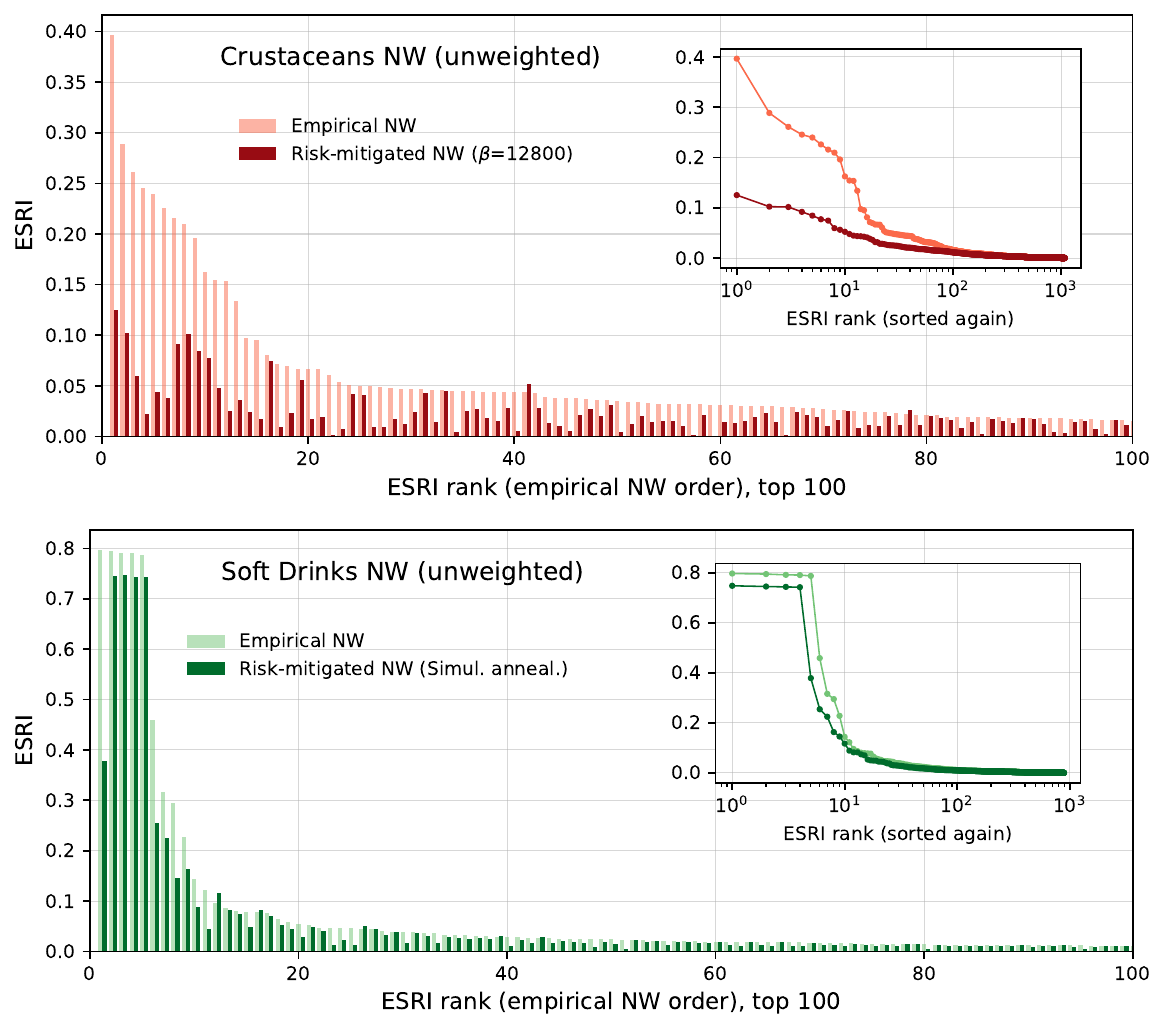}
    \caption{Systemic risk profiles of the crustacean network (top panel) and soft drink network (bottom panel). The networks are considered in their unweigthed version. The light bars report firms' ESRI values in the empirical network, the darker bars the new ESRI values in the rewired networks. We consider the best risk-mitigation results with the largest reduction of $\langle \text{ESRI} \rangle$, fixed $\beta=12800$ and simulated annealing. Firms are ranked by their ESRI in the empirical network along the $x$-axis, displaying only the top 100 riskiest firms. The inset presents the same profiles, with the rewired profile now rank-ordered for direct comparison.}
    \label{fig:ESRI profiles bef/aft 2}
\end{figure}

In Fig.~\ref{fig:ESRI profiles bef/aft 1} and~\ref{fig:ESRI profiles bef/aft 2}, we report for each network the comparison between the ESRI attributes of the 100 riskiest nodes in the empirical data and their new values in the most effective minimization. In the main plot of each subfigure, the firms are ranked by their ESRI values in the empirical network along the $x$-axis, with only the riskiest 100 firms displayed. For each firm, ESRI in the empirical network (lighter color bars) is compared to ESRI in the risk-mitigated network. In the inset, firms are re-ordered by their ESRI values in both the empirical and mitigated networks, allowing for a direct comparison of the distribution profiles. This means that a firm's rank in the empirical network (lighter line) may differ from its rank in the mitigated network (black line).

\FloatBarrier

\FloatBarrier

\begin{figure}
    \centering
    \vspace{5cm}
    \includegraphics[width=410pt]{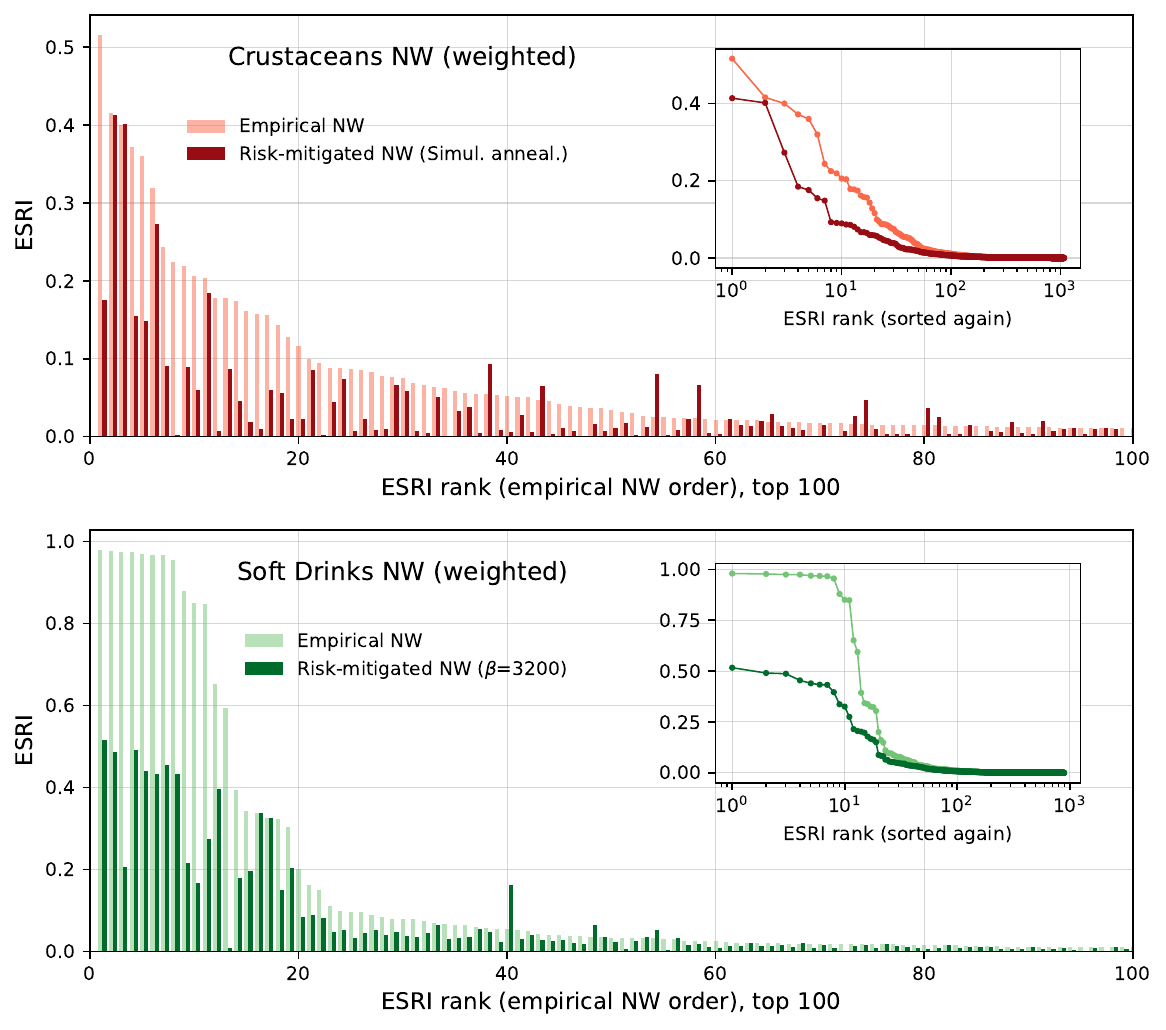}
    \caption{Systemic risk profiles of the crustacean network (top panel) and soft drink network (bottom panel). The networks are considered in their weigthed version. The light bars report firms' ESRI values in the empirical network, the darker bars the new ESRI values in the rewired networks. We consider the best risk-mitigation results with the largest reduction of $\langle \text{ESRI} \rangle$, simulated annealing and fixed $\beta=3200$. Firms are ranked by their ESRI in the empirical network along the $x$-axis, displaying only the top 100 riskiest firms. The inset presents the same profiles, with the rewired profile now rank-ordered for direct comparison.}
    \label{fig:ESRI profiles bef/aft 3}
\end{figure}
\FloatBarrier
\newpage

\subsection{Random exploration of the configuration space ($\beta=0$)}\label{SI:sectBETA0}
We compared the empirical networks with the ones resulting from the $\langle\text{ESRI}\rangle$ minimization considering several network measures, and reported the results in Table 1 in the main text. Some of the measures are different before and after the minimization process. Their different final values may be distinctive characteristics of the more robust network states, or simply an emergence of the configuration model. To understand if any of these changes carries information about the increased robustness, we rewired the networks with the same algorithm without considering the driving Metropolis-Hastings criterion, and so accepting every link rewiring.

\begin{figure}[!h]
    \centering
    \includegraphics[width=500pt]{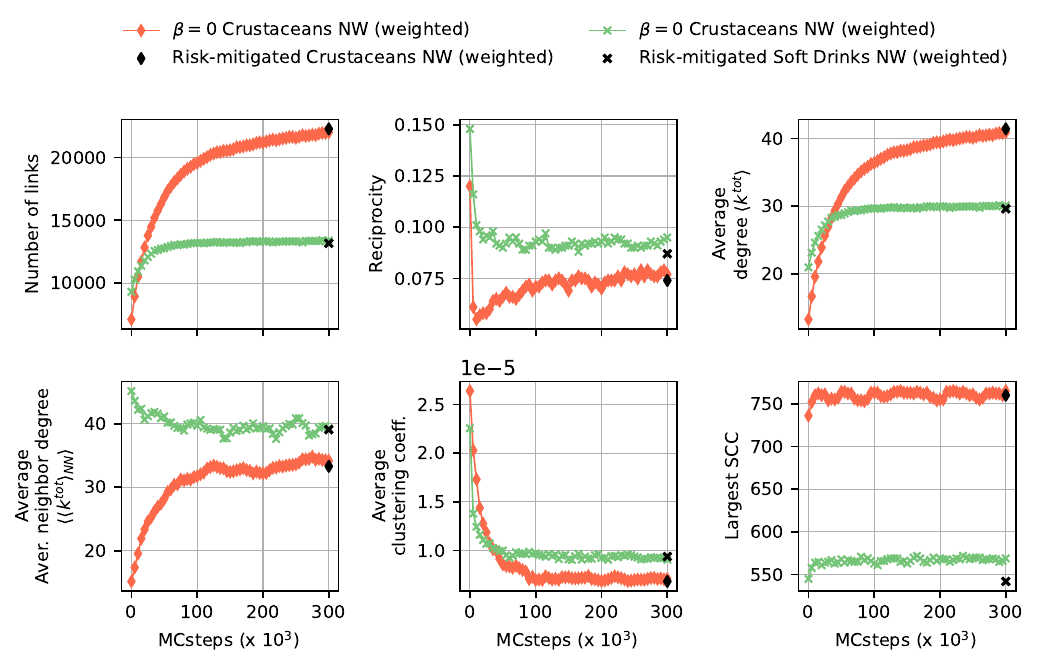}
    \caption{Evolution of several network measures for the configuration model. Link rewiring is always accepted, without evaluating $\Delta\langle\text{ESRI}\rangle$ in the Metropolis-Hastings criterion. In each subfigure, for each weighted network, the trajectory of the measure is compared with the final value of the simulation minimizing $\langle\text{ESRI}\rangle$.}
    \label{fig:weighted B0 comparison c2}
\end{figure}

\begin{figure}
    \centering
    \vspace{5cm}
    \includegraphics[width=500pt]{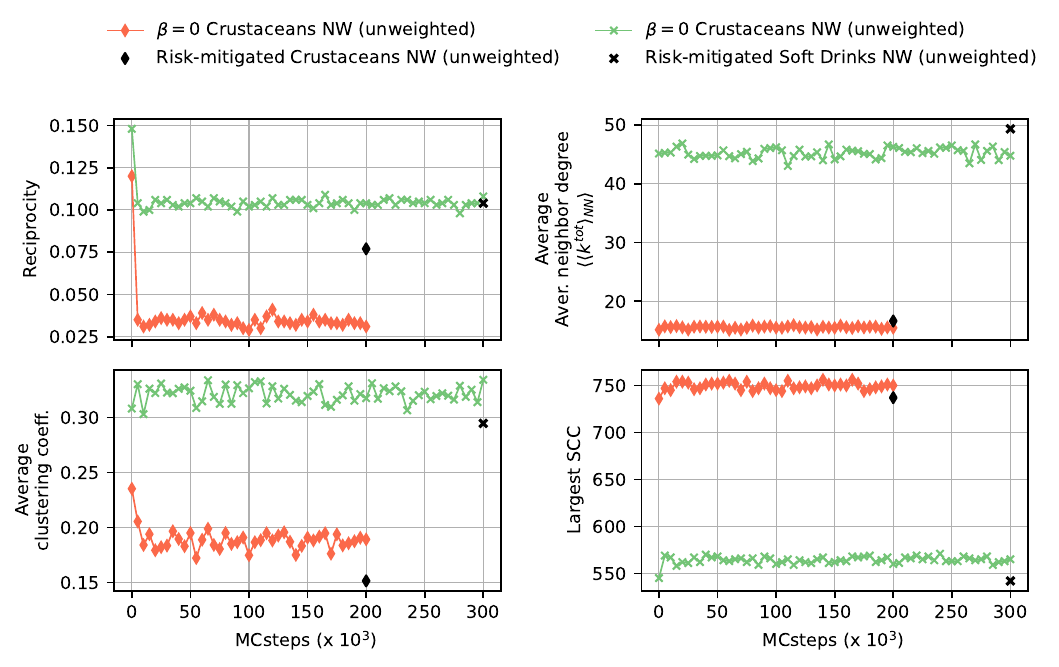}
    \caption{Evolution of several network measures for the configuration model. Link rewiring is always accepted, without evaluating $\Delta\langle\text{ESRI}\rangle$ in the Metropolis-Hastings criterion. In each subfigure, for each unweighted Ecuadorian network, the trajectory of the measure is compared with the final value of the simulation minimizing $\langle\text{ESRI}\rangle$.}
    \label{fig:unweighted B0 comparison c2}
\end{figure}

\begin{figure}
    \centering
    \vspace{5cm}
    \includegraphics[width=500pt]{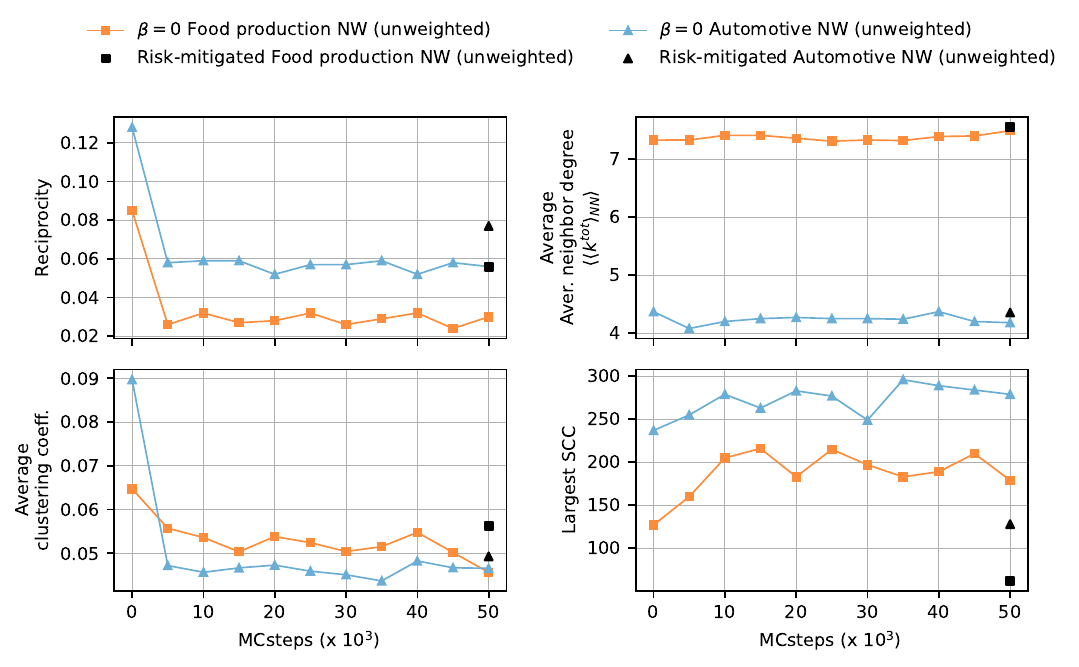}
    \caption{Evolution of several network measures for the configuration model. Link rewiring is always accepted, without evaluating $\Delta\langle\text{ESRI}\rangle$ in the Metropolis-Hastings criterion. In each subfigure, for each unweighted Hungarian network, the trajectory of the measure is compared with the final value of the simulation minimizing $\langle\text{ESRI}\rangle$.}
    \label{fig:unweighted B0 comparison c1}
\end{figure}

For each network measure that has changed with the $\langle\text{ESRI}\rangle$ minimization, we compared the measure's trajectory in the non-driven rewiring simulation, considering the network state every 5000 Monte Carlo steps, and the measure's punctual value in the network with the best $\langle\text{ESRI}\rangle$ minimized result. In the weighted Ecuadorian networks, we considered the number of links $L$, the reciprocity of the links, the average total degree $\langle k^{tot}\rangle$, the mean average neighbors' total degree $\langle\langle k^{tot}\rangle_{NN}\rangle$, the average local clustering coefficient, and the size of the largest strongly connected component. The results are reported in Fig.~\ref{fig:weighted B0 comparison c2}, with one network measure per subfigure and trajectories colored after the considered network. The black points represent the $\langle\text{ESRI}\rangle$ minimized networks. Looking in particular at the first subfigure in Fig.~\ref{fig:weighted B0 comparison c2}, the increased number of links reaches similar values in the driven and non-driven rewiring. This means that $L$ does not explain the decrease in systemic risk in the risk-mitigated network, which informs us that the robustness does not come from just diversifying the number of suppliers or customers. Because the unweighted networks do not change the number of links, we considered only reciprocity, mean average neighbors' total degree $\langle\langle k^{tot}\rangle_{NN}\rangle$, average local clustering coefficient, and size of the largest strongly connected component. Results are plotted in Fig.~\ref{fig:unweighted B0 comparison c1}~(Ecuador) and~\ref{fig:unweighted B0 comparison c2} (Hungary).

\FloatBarrier

\FloatBarrier

\FloatBarrier

\subsection{Firms' connectivity and systemic risk contribution}\label{SI:sectSCATTERPLOTS}
\begin{figure}
    \centering
    \includegraphics[width=320pt]{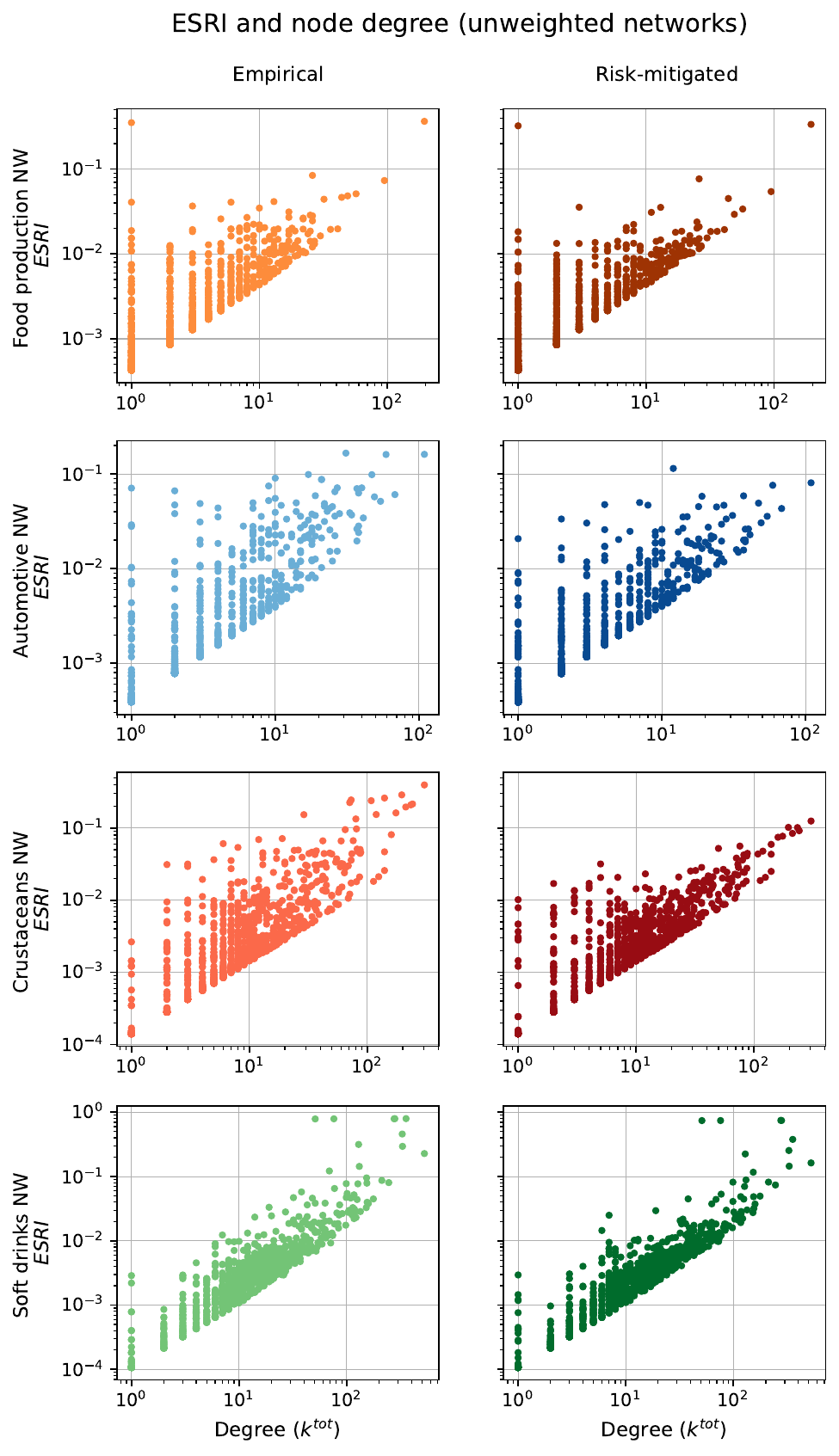}
    \caption{Relationship between firms' Economic Systemic Risk Index (ESRI) and node total degree ($k^{tot}$) across the four unweighted networks. Each row corresponds to a distinct network: Food production, Automotive, Crustaceans, and Soft drinks. The left column shows results for empirical networks, while the right column presents risk-mitigated networks. For unweighted networks, the rewiring algorithm reduces the overall risk magnitudes without affecting node degrees. The plots highlight the positive correlation between node degree and ESRI. A high degree often indicates high systemic importance, but the reverse is not always true. Axes are displayed on logarithmic scales.
    }
    \label{fig:ESRIvsdegree}
\end{figure}

\begin{figure}[t]
    \centering
    \includegraphics[width=320pt]{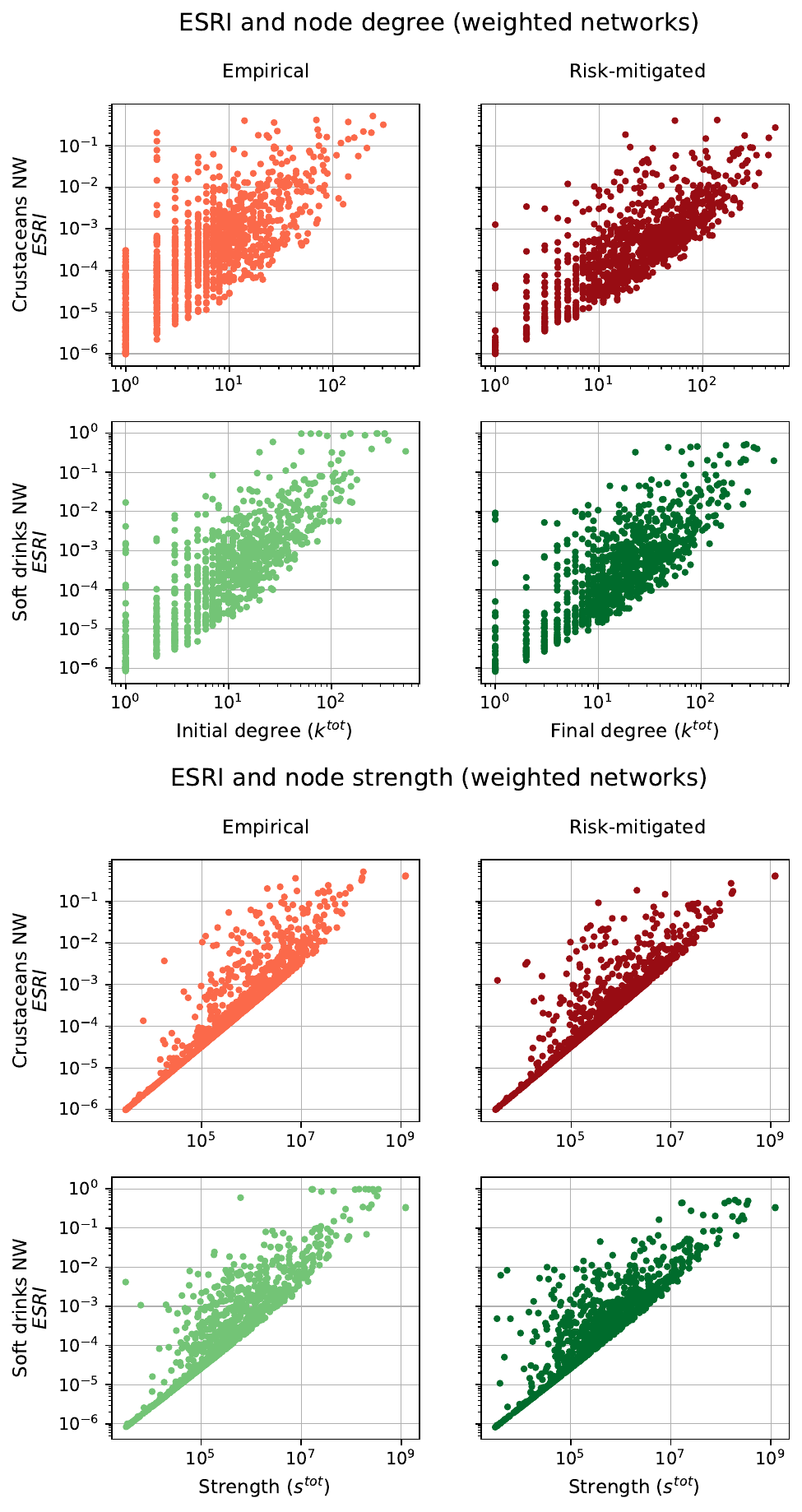}
    \caption{Relationship between firms' Economic Systemic Risk Index (ESRI), node total degree ($k^{tot}$) and total strength ($s^{tot}$) across the two weighted networks. Each row corresponds to a distinct network: Crustaceans, and Soft drinks. The left column shows results for empirical networks, while the right column presents risk-mitigated networks. For weighted networks, the rewiring algorithm reduces the overall risk magnitudes, preserving the strength per sector and occasionally changing node degrees. }
    \label{fig:ESRIvsStrength}
\end{figure}

Since a firm's ESRI is determined by the cascading production failures triggered when its supply orders and output deliveries are halted, it heavily depends on the relevance of the firm's connections to the production of its supply chain neighbors. Consequently, firms with high in, out, or total strength - whether large hubs connected to many firms or those with only a few but highly weighted links - tend to trigger larger cascades and have high ESRI. A key finding of \cite{Diem2022ESRI} is that even small yet essential suppliers to large companies can carry significant systemic risk. 
In unweighted networks, where all links are treated equally, node strength corresponds to node degree.
We analyze the relationship between firms' connectivity and ESRI in our subnetworks and report in Fig.~\ref{fig:ESRIvsdegree} and~\ref{fig:ESRIvsStrength} the scatterplots of node strength or degree against ESRI (in log-log scale). For each row, the subfigure on the left and on the right show the results for the empirical production network and for the best result of the risk-mitigation rewiring. For unweighted networks, the rewiring algorithm preserves firms' degree, but alters their neighborhood. While degree values remain unchanged, ESRI values tend to decrease after rewiring. The correlation between degree and ESRI is evident, and rewiring makes the scatterplot points more concentrated, with a more linear dependence (in log-log scale). For weighted networks, ESRI appears to exhibit a stronger linear dependence on node strength (again, in log-log scale). Unlike in \cite{Diem2022ESRI}, where many small firms (low strength) were found to have high systemic importance in the Hungarian SCN, our analysis does not show this pattern clearly. We attribute this to the economic activity (NACE \textit{groups}) filtering process, which may have excluded these small-size suppliers of essential goods for large companies. However, firms with strengths several orders of magnitude smaller than the largest firms but comparable ESRI values are still present.

Lastly, while a firm's node degree remains unchanged between weighted and unweighted network representations, its ESRI can vary due to the different technological coefficients for firm production. The availability of weight information increases the level of detail in modeling the relevance of a firm for its suppliers and customers, and thus the potential cascade of failures through its network neighbors.

\FloatBarrier

\FloatBarrier

\subsection{Weighted networks rewiring and relationship between change in node degree and ESRI}\label{SI:sectdeltaKdeltaESRI}
The rewiring algorithm can alter the number of links $L$ in the network, either when the two links selected for swapping have significantly different weights (see Fig. 2 in the main text) or when the cross-links already exist (see Fig.~\ref{fig:pathological_cases} \textbf{a)}). Since, in practice, the rewiring process in weighted networks substantially increases the number of links (see Tab.1 in the main text), we investigate whether the observed systemic risk mitigation results from diversification—i.e., firms gaining more suppliers and customers. The simulation results at $\beta=0$ already demonstrate that simply increasing $L$ is not a sufficient condition for risk mitigation. At this parameter value, $L$ is comparable to that in the best risk-mitigation configurations, yet the average systemic risk $\langle\text{ESRI}\rangle$ remains at empirical levels. In this section, we examine to what extent increasing $L$ is a necessary condition for risk mitigation.

\begin{figure}
    \centering
    \vspace{1cm}
    \includegraphics[width=500pt]{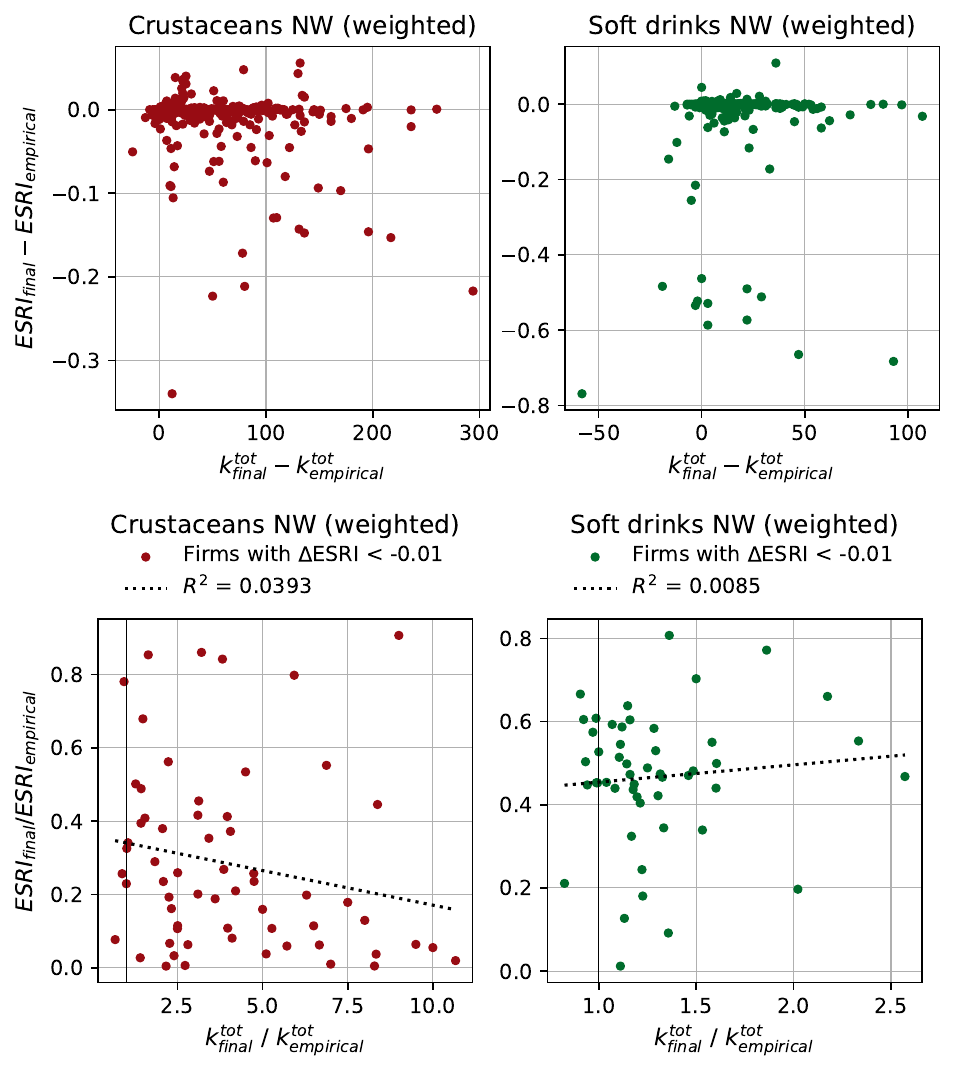}
    \caption{Scatterplot $\Delta$ $k$ against $\Delta$ ESRI (all nodes), and scatterplot with the relative $k$ change against relative ESRI change focusing the firms that have a significant $\Delta$ ESRI.}
    \label{fig:deltakdeltaESRI}
\end{figure}

We focus on the best risk-mitigation results in the weighted crustaceans and soft drinks production networks, comparing the changes in node degree $k^{tot}$ and ESRI in the networks before and after the rewiring simulation. The top panel of Fig.~\ref{fig:deltakdeltaESRI} presents scatterplots showing how each node's $k^{tot}$ and ESRI value changed, in terms of absolute $\Delta$. While increasing $k^{tot}$ generally reduces ESRI (evidenced by more points in the bottom right than the bottom left), many firms decrease their ESRI without a significant change in $k^{tot}$. Notably, in the soft drinks network in particular, some firms even concentrate their transactions on fewer partners ($\Delta \text{ESRI} < 0$, $\Delta k^{tot} < 0$). This indicates that diversification is not strictly necessary for individual firms to reduce systemic risk.

On a broader scale, the network undergoes a structural reorganization where increasing the number of connections for some nodes can reduce—or even increase—ESRI for others. However, relatively few firms experience an increase in ESRI (region at $\Delta\text{ESRI} > 0$), which aligns with our observations from Fig. 4 in the main text: while a few outliers increase their ESRI, the majority of the firms experience a reduction. One possible scenario could have been that the riskiest firms redistribute their systemic importance across smaller firms, leading to a lower average ESRI at the cost of shifting the tail of the ESRI distribution upward. However, this does not appear to be the case.

Because node strength and degree strongly correlate with systemic importance (see Fig.~\ref{fig:ESRIvsStrength}) and follow a fat-tailed distribution, not all firms play a major role in systemic risk mitigation. This results in a dense cluster of points around $\Delta\text{ESRI} = 0$. We focus on firms that significantly contributed to risk mitigation by reducing their ESRI by at least 0.01, as the self-organization of high-ESRI firms towards a more resilient network may follow different dynamics. The bottom panel of Fig.~\ref{fig:deltakdeltaESRI} shows the relative changes in degree and ESRI. A vertical line distinguishes regions where firms had relatively lower ($<1$) or higher ($>1$) degrees after rewiring. Even for these larger contributors, the pattern seems the same: increasing the degree generally reduces ESRI, but some firms achieved a 50\% ESRI reduction despite a lower degree after rewiring (soft drinks network).

\FloatBarrier
\end{document}